\documentclass[fleqn,usenatbib]{mnras}

\usepackage{newtxtext,newtxmath}
\usepackage[T1]{fontenc}

\usepackage{graphicx}
\graphicspath{{./}{figures/}}
\usepackage{subcaption}
\usepackage{color}
\usepackage{amsmath}
\usepackage{verbatim}
\usepackage{hyperref}
\usepackage[noabbrev,capitalize]{cleveref}
\usepackage{natbib}
\usepackage[mathscr]{euscript}
\usepackage{xspace}
\usepackage[normalem]{ulem}
\usepackage{deluxetable}

\makeatletter
\usepackage{etoolbox}
\patchcmd\H@refstepcounter{\protected@edef}{\protected@xdef}{}{}
\makeatother

\newcommand{\sn}{SN~Ia\xspace}
\newcommand{\sne}{SNe~Ia\xspace}

\newcommand{\absmag}{M\ensuremath{'}\xspace}  
\newcommand{\pplus}{Pantheon+\xspace}
\newcommand{\true}[1]{\ensuremath{{#1}_{\mathrm{true}}}}

\newcommand{\lcdm}{$\Lambda$CDM\xspace}     
\newcommand{\h}{\ensuremath{\text{H}_0}\xspace}

\newcommand{\rv}{\ensuremath{R_{V}}\xspace}

\newcommand{\unity}{UNITY\xspace}
\newcommand{\linmix}{\textsc{LINMIX}\xspace}
\newcommand{\salt}{\textsc{SALT2}\xspace}

\newcommand{\co}{\ensuremath{c}\xspace}

\newcommand{\un}[1]{~\text{#1}\xspace}  

\newcommand{\bhm}[1][]{BHM{#1}\xspace}

\newcommand{\Npplus}{1550\xspace}
\newcommand{\Ntotal}{1686\xspace}
\newcommand{\Nmore}{57\xspace}
\newcommand{\Nhighly}{51\xspace} 
\newcommand{\Nextreamly}{6\xspace} 
\newcommand{\NnewCalibrators}{10\xspace}
\newcommand{\NnewCalibratorsAll}{17\xspace}

\newcommand{\cmax}{1.61} 

\newcommand{\betaLINMIX}{\ensuremath{2.98 \pm 0.04}\xspace}
\newcommand{\rvcosmo}{\ensuremath{3.01 \pm 0.05}\xspace}
\newcommand{\rvcosmoMedian}{\ensuremath{3.01}\xspace}
\newcommand{\rvred}{\ensuremath{2.94 \pm 0.12}\xspace}
\newcommand{\rvredMedian}{\ensuremath{2.94}\xspace}
\newcommand{\rvdelta}{\ensuremath{-0.07 \pm 0.11}\xspace}

\title[Highly Reddened Type Ia Supernovae]{Constraining R$_V$ Variation Using Highly Reddened Type Ia Supernovae from the Pantheon+ Sample}

\author[B. M. Rose, et al.]{
B. M. Rose,$^{1}$\thanks{E-mail: benjamin.rose@duke.edu}
B. Popovic,$^{1}$
D. Scolnic$^{1}$
and D. Brout$^{2}$\thanks{NASA Einstein Fellow}
\\
$^{1}$Department of Physics, Duke University Durham, NC 27708, USA\\
$^{2}$Center for Astrophysics, Harvard \& Smithsonian, 60 Garden Street, Cambridge, MA 02138, USA
}
\date{Accepted 2022 August 31. Received 2022 August 31; in original form 2022 June 20}
\pubyear{2022}

\begin{document}
\label{firstpage}
\pagerange{\pageref{firstpage}--\pageref{lastpage}}
\maketitle

\begin{abstract}
Type Ia supernovae (\sne) are powerful tools for measuring the expansion history of the universe, but the impact of dust around \sne remains unknown and is a critical systematic uncertainty. One way to improve our empirical description of dust is to analyse highly reddened \sne ($E(B-V)>0.4$, roughly equivalent to the fitted SALT2 light-curve parameter $c>0.3$).
With the recently released Pantheon+ sample,
there are \Nmore \sne that were removed because of their high colour alone (with colours up to $c=\cmax$), which can provide enormous leverage on understanding line-of-sight $R_V$.
Previous studies have claimed that \rv decreases with redder colour, though it is unclear if this is due to limited statistics, selection effects, or 
an alternative explanation.
To test this claim, we fit two separate colour-luminosity relationships, one for the main cosmological sample ($c<0.3$) and one for highly reddened ($c>0.3$) \sne.
We find the change in the colour-luminosity coefficient to be consistent with zero.
Additionally, we compare the data to simulations with different colour models, 
and find that the data prefers a model with a flat dependence of \rv on colour over a declining dependence. Finally, our results strongly support that line-of-sight $R_V$ to \sne is not a single value, but forms a distribution.
\end{abstract}


\section{Introduction}\label{intro}

\defcitealias{Guy2010}{G10}
\defcitealias{Chotard2011}{C11}
\defcitealias{Mandel2011}{M11}
\defcitealias{Brout2021}{BS21}
\defcitealias{Popovic2022}{P22}

Type Ia supernovae (\sne) were used to discover the acceleration of the universe \citep{Riess1998,Perlmutter1999}, and are now being used to make some of the most precise measurements of the equation-of-state of dark energy \citep[$w$, e.g.,][]{Garnavich1998b,Suzuki2012,Betoule2014,Scolnic2018,DESCollaboration2019,Brout2022b} and the current expansion rate parameterized by the Hubble constant \citep[\h, e.g.,][]{Freedman2019,Riess2022}. However, recent analyses have shown that one path for improvement of these measurements depends on a better understanding of the impact of dust around \sne and how it relates to the scatter found in standardized \sne brightnesses \citep{Brout2021,Popovic2022}. 

One way to gain leverage in characterizing the role of dust is to analyse highly-reddened \sne ($E(B-V) > 0.4$), 
where intrinsic colour variations of the \sn have a marginal effect compared to extragalactic dust \citep{Phillips2013,Scolnic2016}. 
For cosmological analyses with the \sn light-curve fitter SALT2 \citep{Guy2007,Guy2010}, highly-reddened \sn are cut \citep[$-0.3<c<0.3$ with $c \sim E(B-V) - 0.1$, e.g.,][]{Betoule2014,Scolnic2018,Brout2022b}.
These \sne are 
not used in the training of the spectral model itself \citep{Betoule2014,Taylor2021,Brout2022a}, but it has not been shown that they cannot be used.
In this analysis, we investigate these highly-reddened \sne to constrain the evolution and variability  of reddening.

Dust properties are often parameterized with the total-to-selective extinction parameter, \rv. \rv is defined as $A_V /(A_B - A_V)$, where $A_V$ is the extinction in the V band ($\lambda \sim 5500 \un{\AA}$), and $A_B$ is the extinction in the B band ($\lambda \sim 4400 \un{\AA}$). \rv is known to vary for different sizes and composition of dust grains. The diffuse interstellar medium of the Milky Way galaxy has an average \rv of $\sim$3.1 \citep[e.g.,][]{Schultz1975,DeMarchi2021}.

\sn studies typically find \rv values lower than the Milky Way average \citep[e.g.,][]{Jha2007,Goobar2008,Burns2014}. These measurements are with large light-curve samples, however, with a limited colour range ($E(B-V)<0.4$).
There are some \rv measurements with spectrophotometric data \citep[e.g.,][]{Krisciunas2006,Elias-Rosa2006,Phillips2013,Amanullah2015,Cikota2016}. These also typically measure low \rv, but these are mostly from highly reddened \sne, where it is easier to constrain \rv. It is uncertain whether \rv evolves with colour or if these results are due to selection effects.

One of the largest studies of \rv for highly-reddened objects was in \citet[][hereafter \citetalias{Mandel2011}]{Mandel2011}. They look at 127 \sne, with 5 having $E(B-V)>0.5$. 
They found that \sne with a lower dust extinction, $A_V < 0.4 \un{mag}$, have a relationship between colour and absolute luminosity consistent with an $\rv \sim 2.5\text{--}2.9$, while at high extinctions ($A_V > 1 \un{mag}$) low values of $\rv < 2$ were favoured. 
However, this data set is from ``targeted'' \sn surveys where the observations target larger brighter galaxies, possibly biasing the data set's colour demographics.
Contrary to \citetalias{Mandel2011}, \citet{Gonzalez-Gaitan2021} finds that \rv grows with \sn colour, though with a limited colour range to $-0.3<c<0.3$. 
These conflicting results indicate the complexity of the relationship between \rv and \sn colour.

It is also unclear if there is a distribution of \rv values affecting measurements of \sne.
\citet{Amanullah2015} spectrophotometrically measured two \sne (SN 2012cu, SN 20214J) with $E(B-V) \gtrsim 1$, one with an $\rv = 1.4 \pm 0.1$ and the other with an $\rv = 2.8 \pm 0.1$, respectively. This would indicate that \sne of the same observed colour can have a range of \rv values.
Similarly, \citet{Huang2017} measured a \sn with an $\rv = 2.95 \pm 0.08$ with a $c = 0.90 \pm 0.03$, contrary to the trend with colour seen in \citetalias{Mandel2011}.
This variable \rv for a given colour is one of the main ideas of \citet{Brout2021}.
\citet[][hereafter \citetalias{Brout2021}]{Brout2021} assumes an intrinsic \sn colour and colour-luminosity relationship characterized by Gaussian distributions that are reddened by independent column-density extinction and selective extinction ($E(B-V)$ and \rv respectively).
A distribution of line-of-sight $R_V$ values is inferred in the Dark Energy Survey's analysis of \sne in redMaGiC galaxies \citep{Chen2022} and it is shown in \citet{Meldorf2022} that $R_V$ determined from \sne light curves appear to correlate with $R_V$ values of the \sne hosts.

In this work, we investigate if highly reddened \sne have a consistent colour-luminosity relationship when comparing to the colour-luminosity relationship of the traditional cosmological sample.
Since highly reddened \sn are fainter, selection effects will be more prominent. Selection effects, along with a variable \rv like the one described in
\citetalias{Brout2021}, would result in
observed \rv values being smaller than the true population mean, because the lowest \rv values are brighter and therefore easier to observe. This combination could lead to a bias in measured \rv values.

First, we present the \pplus data set, including the highly reddened sub-sample (\cref{sec:data}).
We then discuss several empirical  colour luminosity relationships as well as the ways we forward modelled the \pplus selection effects in \cref{sec:method}.
We present our results in \cref{sec:results}. 
Finally, we discuss the implications of these results by comparing them to previous measurements and present a specific details about how these results affect future cosmological analyses (\cref{sec:discussion}).

\section{Data}\label{sec:data}

For this analysis, we use the \pplus data set \citep[the latest aggregated data set of spectroscopically confirmed \sne]{Scolnic2022}, but without cutting the highly reddened \sne. The main data set contains \Npplus individual \sn (with 1701 total light curves) extending to a redshift of $z=$~2.26. The photometry of the light curves are all corrected for Milky Way reddening.
\pplus is released as a series of papers. The redshifts and peculiar velocities of the \sne are described in \citet{Carr2021}. A comprehensive analysis of redshift systematics are presented in \citet{Peterson2021}. The cross-calibration of the different photometric systems is presented in \citet{Brout2022a}, and calibration-related systematic uncertainty limits were determined in \citet{Brownsberger2021}. 
These works culminate in measurements of the Hubble constant and dark energy equation-of-state \citep[respectively]{Riess2022, Brout2022b}.
The full data set is available at \url{ pantheonplussh0es.github.io}.

\subsection{\salt}\label{sec:salt}

In order to use \sne as standardizable candles, the \pplus data set 
provides fit parameters from \salt \citep{Guy2007,Guy2010,Mosher2014} light-curve fits
for each \sne using the latest retraining and cross-survey calibration \citep{Taylor2021,Brout2022a}.
The code fits light curves in broad band photometry using spectral energy distribution (SED) training.
The model separates out the temporal and colour variations of \sn. The \salt model can be approximately described as
\begin{equation}\label{eqn:saltcolour}
    x_0 (M_0(t,\lambda) + x_1 M_1 (t,\lambda))\times e^{c~\mathrm{CL}(\lambda)}~~.
\end{equation}
Here, $x_0$ is the overall flux normalization factor, $M_0$ is the mean spectral template, and $t$ is the rest-frame day since maximum luminosity in $B$-band. Deviations from the mean are modelled by $M_1$, $x_1$ characterizes the \sn light-curve shape.
CL$(\lambda)$ represents the average colour-correction law.
The colour law obtained when training \salt is very close to \citet{Cardelli1989a} with \rv = 3.1, however, there is a significant divergence between the two colour laws at wavelengths of $<$4000~\AA. Specific deviations between the two colour laws are discussed in the original \salt paper and the recent retrainings \citep[respectively]{Guy2007,Taylor2021}.
$c$ is the fitted colour of the \sn light curve, and its definition is dependent on the exact CL$(\lambda)$ resulting from the model training, but can be approximated as $E(B-V) - 0.1$.

To obtain a standardized absolute magnitude for use as a cosmological distance indicator, the \salt parameters $x_{1}$ and $c$ are used in the Tripp equation \citep{Tripp1998}. The standardized absolute magnitude (M) of each \sn is
\begin{equation}\label{eqn:absmag-full}
    \mathrm{M} = m_B - \mu(z,~\mathrm{cosmology}) + \alpha x_{1} - \beta c + \delta
\end{equation}
where $m_B$ is the apparent magnitude of a single \sn in the rest-frame $B$-band, $\mu$ is its theoretical distance modulus dependent on cosmology, $\alpha$ and $\beta$ are the global, linear standardization coefficients for the \sn light-curve shape and colour parameters, respectively, $\delta$ represents several additional small corrections like the host galaxy ``mass step'' or for observational biases \citep[e.g.][]{Sullivan2010,Smith2020,Popovic2021,Popovic2022}. Finally, we define $\mathrm{M}$ to vary with each \sn due to intrinsic scatter and measurement noise.

\subsection{Highly Reddened \sne}

\begin{figure}
    \centering
    \includegraphics[width=.99\columnwidth]{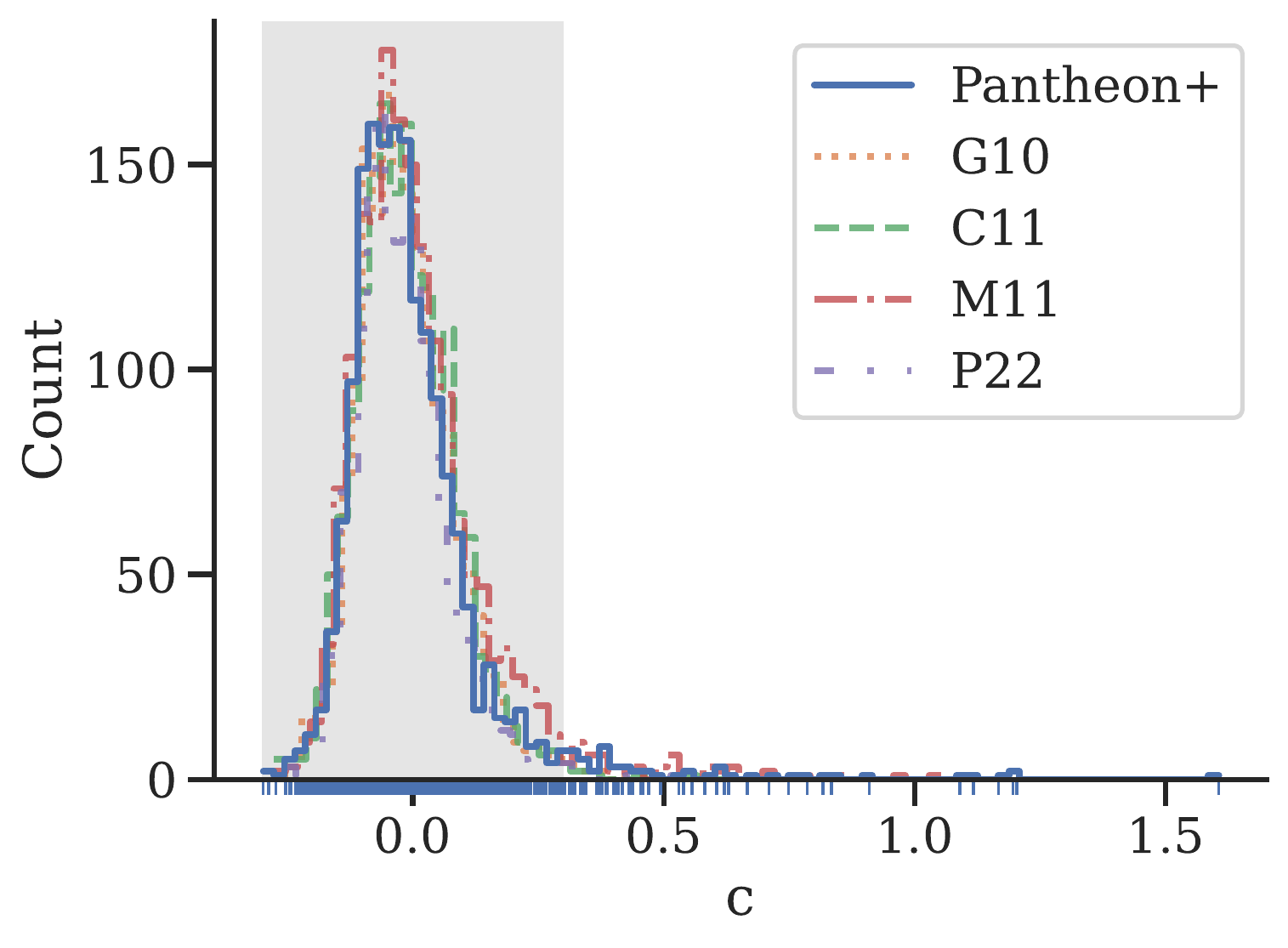}
    \caption{
    \salt colour distribution for our final data set ($N = \Ntotal$). We are still dominated by cosmological \sne (gray shaded region, $-0.3 < c < 0.3$). This sample has \Nmore more \sne than \pplus: \Nhighly have $0.3 < c \leq 1.0$ and \Nextreamly have $c > 1.0$.
    The mean of the cosmological sample and the full sample are  $c=-0.021$ and $c=-0.002$ respectively. The \salt training set has a mean color of 0, by definition.
    Simulations of the \pplus catalogue using four scatter models (\citetalias{Guy2010}, \citetalias{Chotard2011}, \citetalias{Mandel2011}, and \citetalias{Popovic2022}) of \sne are also shown.
    }
    \label{fig:colour}
\end{figure}

For this analysis, we remove the requirement that \sn have $c \le 0.3$. We do, however, keep all the other quality cuts\footnote{The \pplus quality cuts used in this analysis are $P_{\mathrm{fit}} > 0.0$,
$\sigma(x_1)< 1.5$,
$\sigma(\mathrm{pkmjd}) < 2$,
$-0.3 < c$,
$-3 < x_1 < 3$,
$E(B-V)_{\mathrm{MW}} < 0.20 \un{mag}$, and
$T_{\mathrm{rest}} < 5$.} as defined in \citet[]{Scolnic2022}.
We simplify the \texttt{FITPROB} cut from \pplus, by only ensuring that \texttt{FITPROB} is non-zero instead of using survey specific cuts.
We present the distribution of \salt colour in \cref{fig:colour}.
There are \Nmore more \sne than in \pplus: \Ntotal in total, \Nhighly have $0.3 < c \leq 1.0$ and \Nextreamly have $c > 1.0$. Beyond these \Nmore \sne, \pplus also cuts \sne at the bias correction stage, something not addressed in this work.
A similar distribution is seen in the volume limited sample from the Zwicky Transient Facility Bright Transient Survey \citep{Sharon2022}.

\begin{figure}
    \centering
    \begin{subfigure}[b]{0.47\textwidth}
         \centering
         \includegraphics[width=\textwidth]{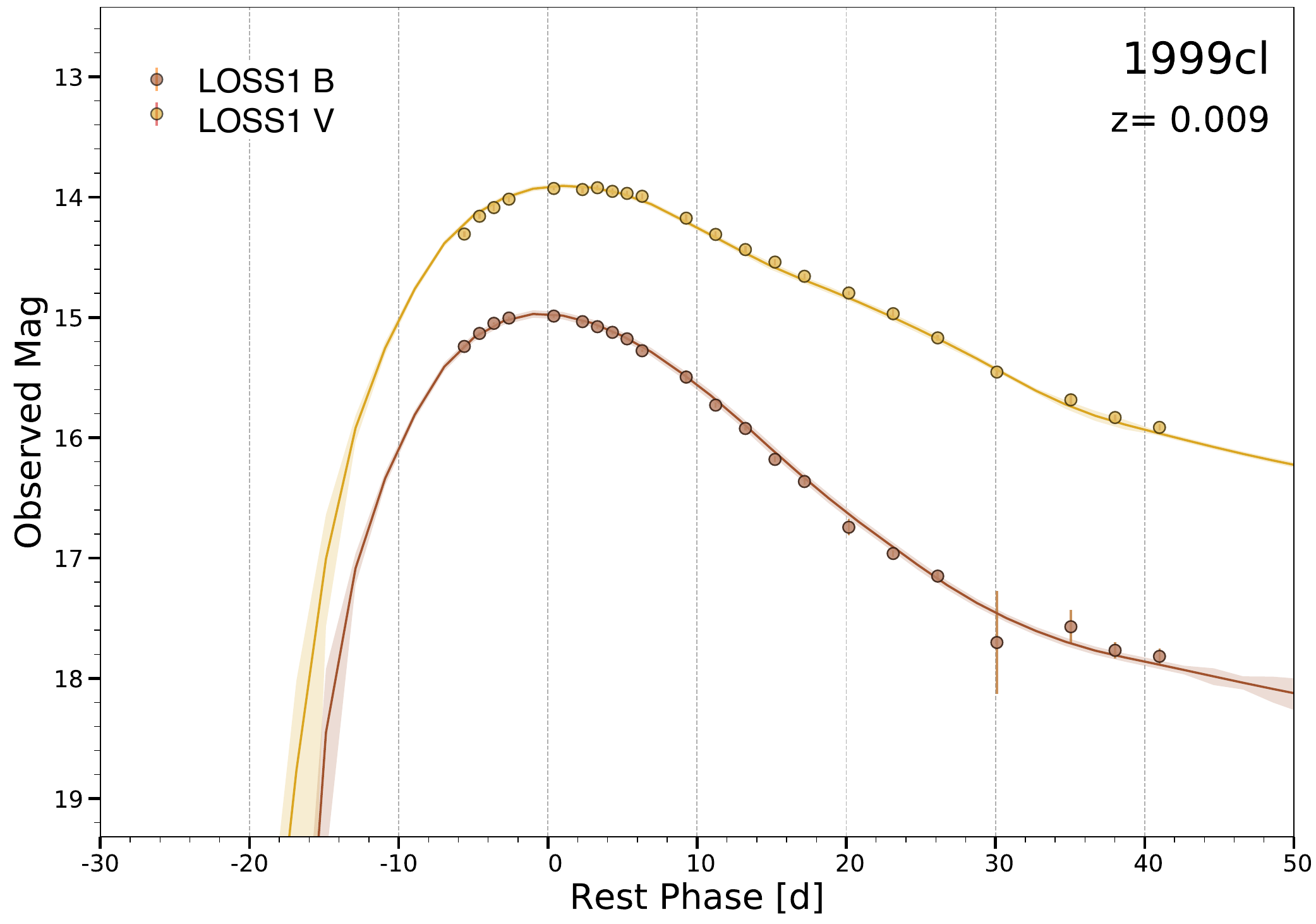}
     \end{subfigure}
    \\
     \begin{subfigure}[b]{0.47\textwidth}
         \centering
         \includegraphics[width=\textwidth]{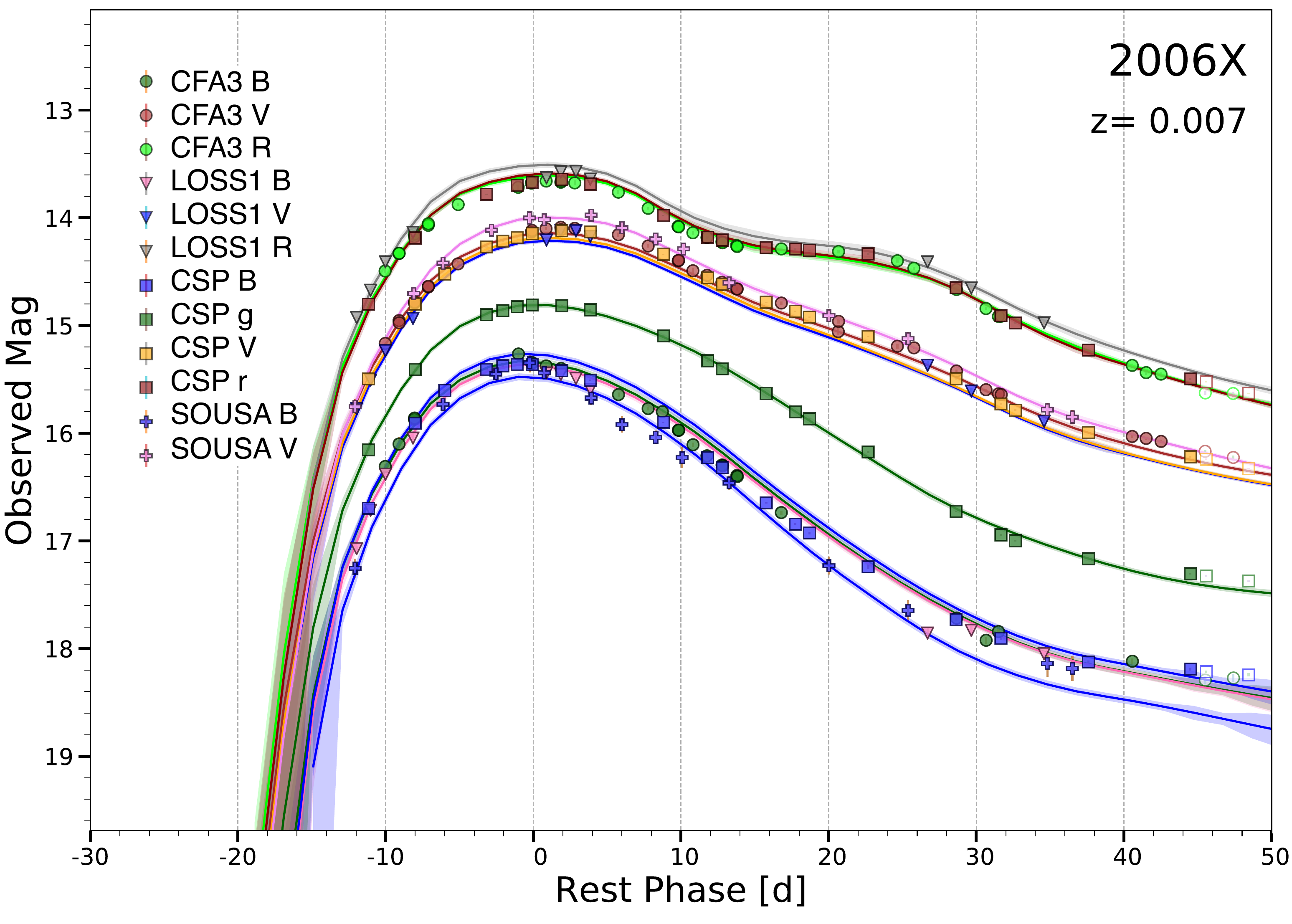}
     \end{subfigure}
    \\
    \vspace{3pt}
     \begin{subfigure}[b]{0.47\textwidth}
         \centering
         \includegraphics[width=\textwidth]{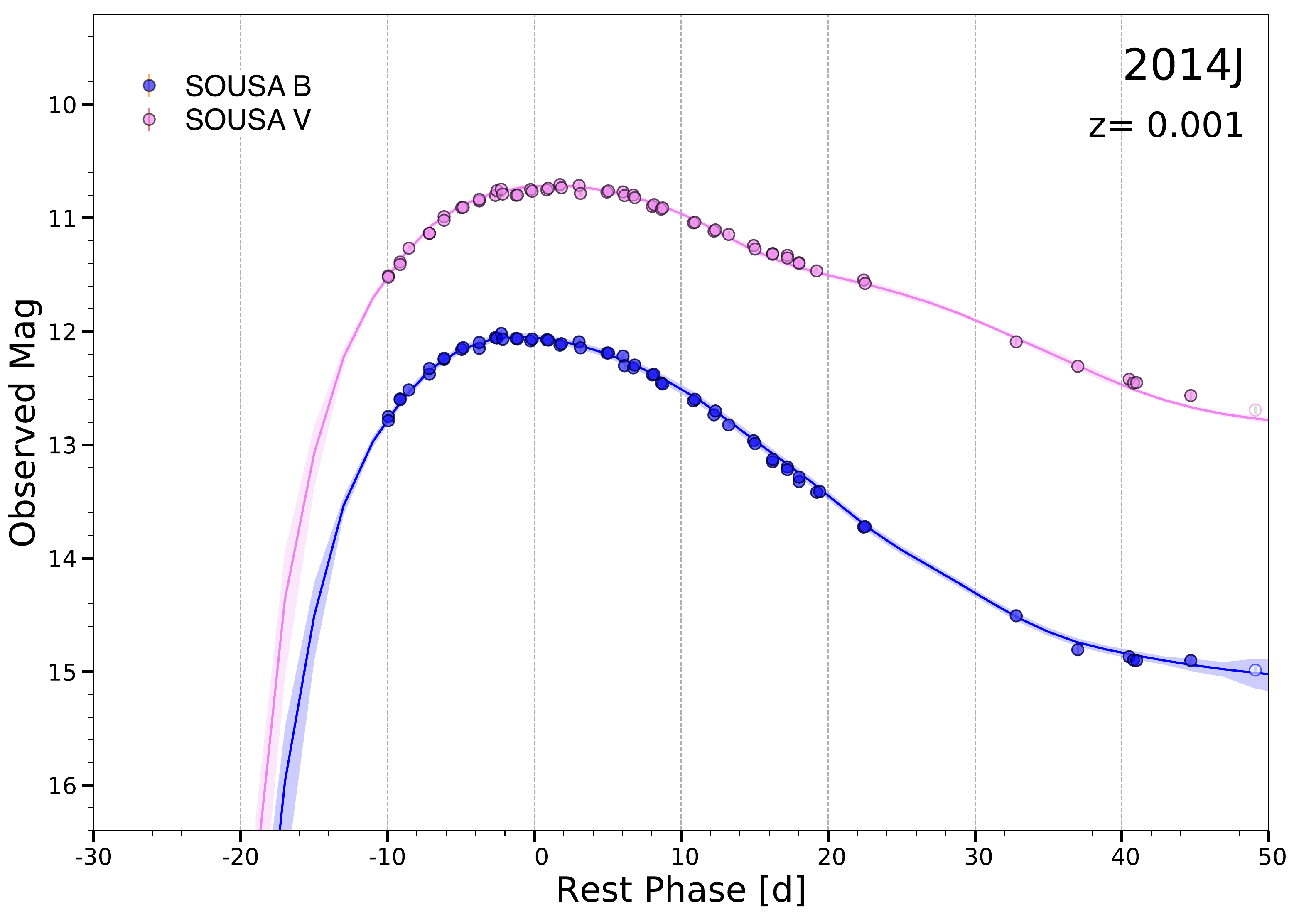}
     \end{subfigure}
    \caption{\salt light-curve fits for three representative extremely reddened \sn ($c>1$). The \salt model is not trained on \sn with $c > 0.3$.
    However, SN1999cl (\textit{left}), SN2006X (\textit{center}), and SN2014J (\textit{right}) have excellent \salt fits, even with \co values of $1.09 \pm 0.04$, $1.19 \pm 0.03$, $1.20 \pm 0.05$ respectively.
    }
    \label{fig:salt}
\end{figure}

The full \pplus data set includes \sn observed from multiple telescopes. These are recorded, initially, as separate events to allow for cross-survey calibrations. However, for this analysis, they are combined into a single object. 
If the same \sn is observed by multiple telescopes, we set the colour and light-curve shape to be the mean of the observations. For the uncertainties on the average parameter, we take the maximum of the individual uncertainties or the range between the individual fits, whichever is greater. This ensures our final uncertainty encompasses all individual point estimates. In the \pplus data release, there are two \sne with data from four surveys (SN2006X and SN2007af). For these objects, we allow ourselves to reject one survey if the light-curve fits disagrees with the other three surveys.
Following this reasoning, we remove the SWIFT data on SN2006X.
We do not find similarly obvious discrepancies for \sn with data from three surveys.

We inspect the quality of the fits for several highly-reddened \sne because of the concern that the light-curve model is not valid at this colour range.
In \cref{fig:salt}, we show three representative light curves of extremely reddened \sn: SN2014J ($c=1.20$), SN1999cl ($c=1.09$), and SN2006X ($c=1.17$). These \sn were selected for presentation prior to seeing their light curves. They all have excellent \salt fits, even with \co values $>$1.

We looked at the percent of \sne that pass the light-curve \texttt{FITPROB} cut described in \citet{Scolnic2022}. In general, red \sne ($c>0.3$) have a lower percent of passing, from a $\sim$90\% to a $\sim$80\% pass rate. The highly reddened \sne have some colour regions with a low percent of passing this quality cut ($\sim$60\% at $c\sim 0.5$) but other regions where 70\%--100\% pass ($0.6 < c < 1.4$),
implying the trustworthiness of light-curve fits resulting in $c > 0.3$.
The increase in fit quality after $c=0.5$ is likely do to stochastic effects since there are only $\sim$4 \sne per $\Delta c=0.1$ bin at this point in the population distribution (\cref{fig:colour}).

\begin{figure*}
    \centering
    \begin{subfigure}[b]{0.32\textwidth}
         \centering
         \includegraphics[width=\textwidth]{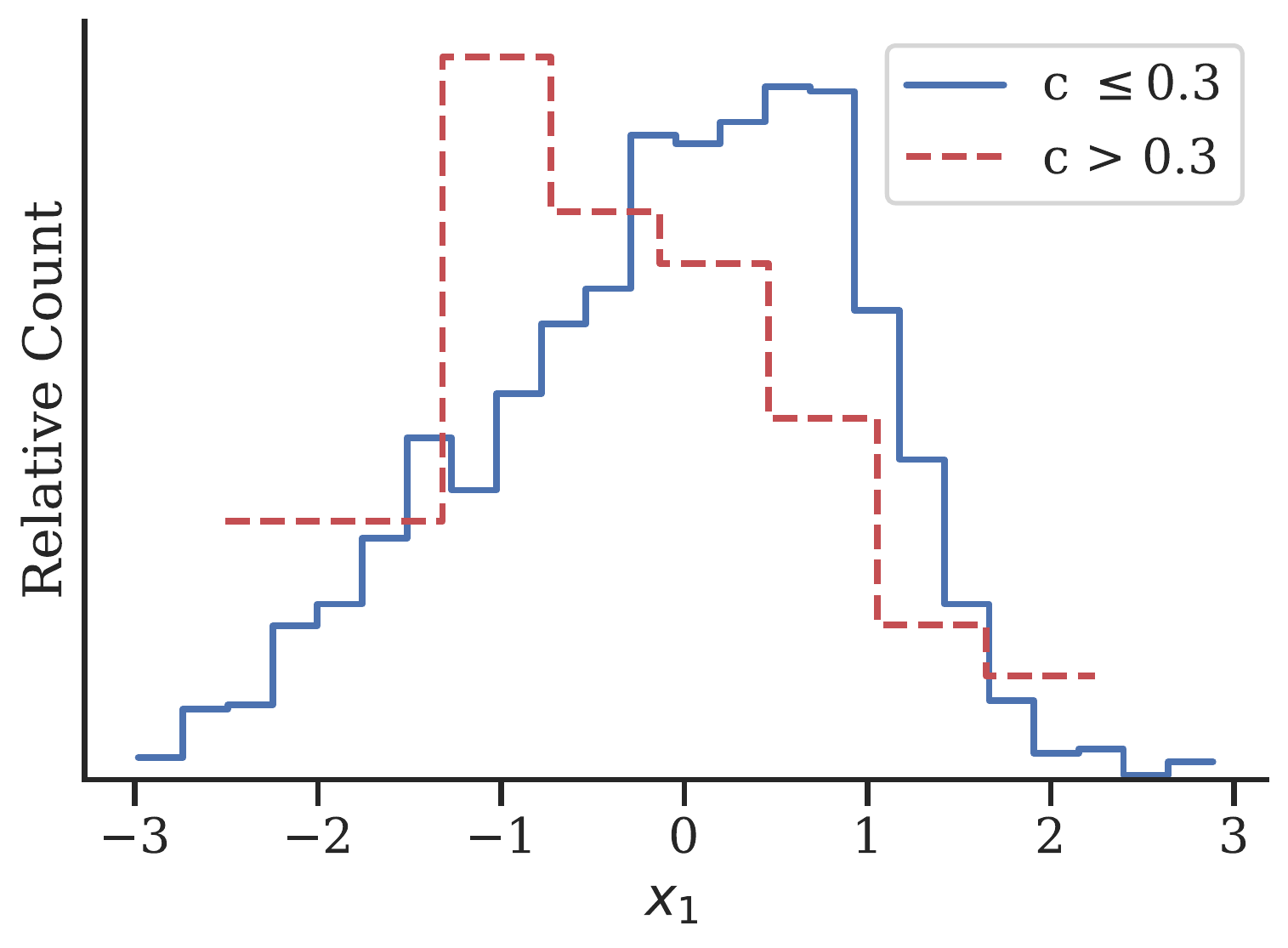}
     \end{subfigure}
     \hfill
     \begin{subfigure}[b]{0.32\textwidth}
         \centering
         \includegraphics[width=\textwidth]{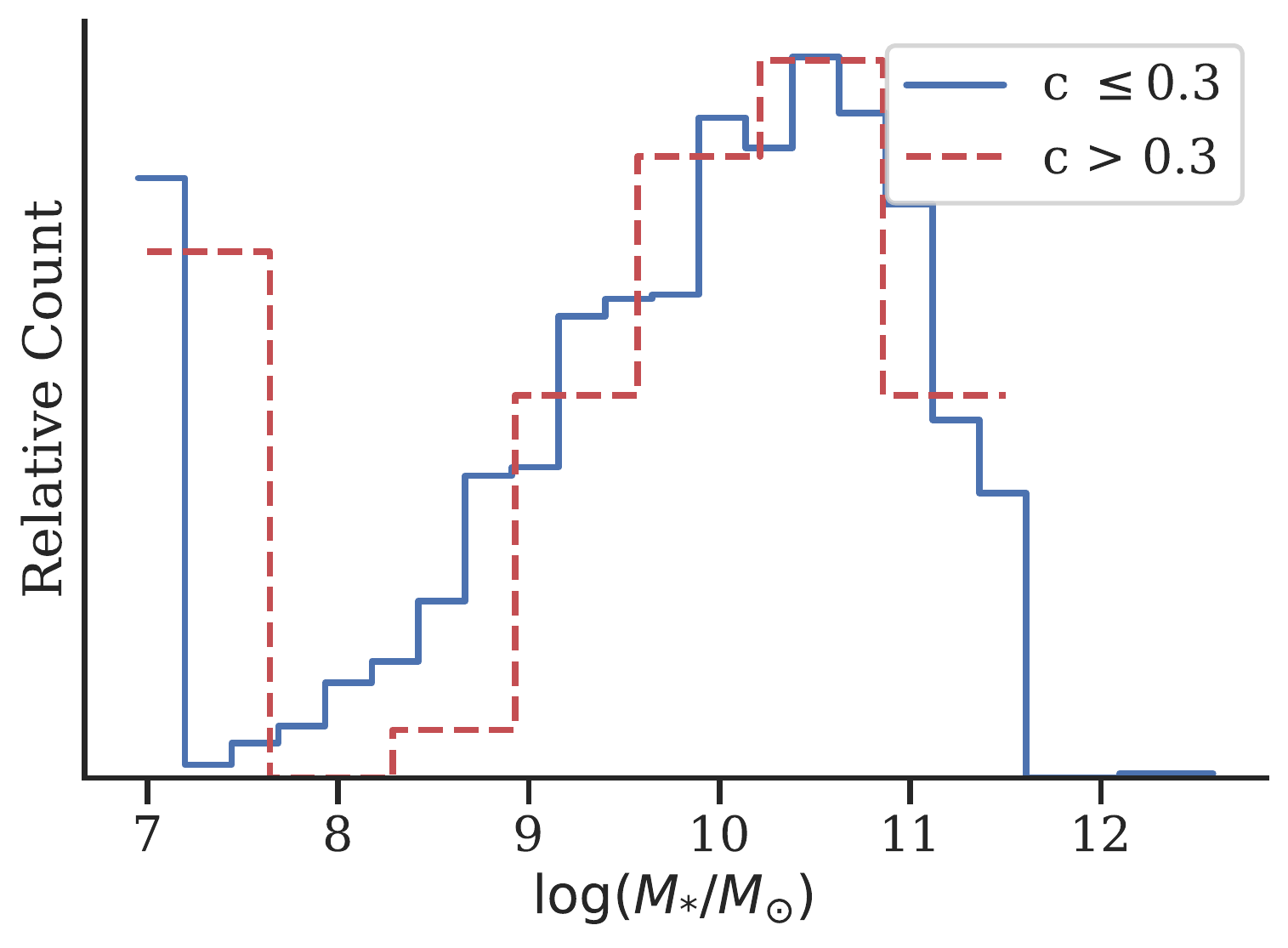}
     \end{subfigure}
     \hfill
     \begin{subfigure}[b]{0.32\textwidth}
         \centering
         \includegraphics[width=\textwidth]{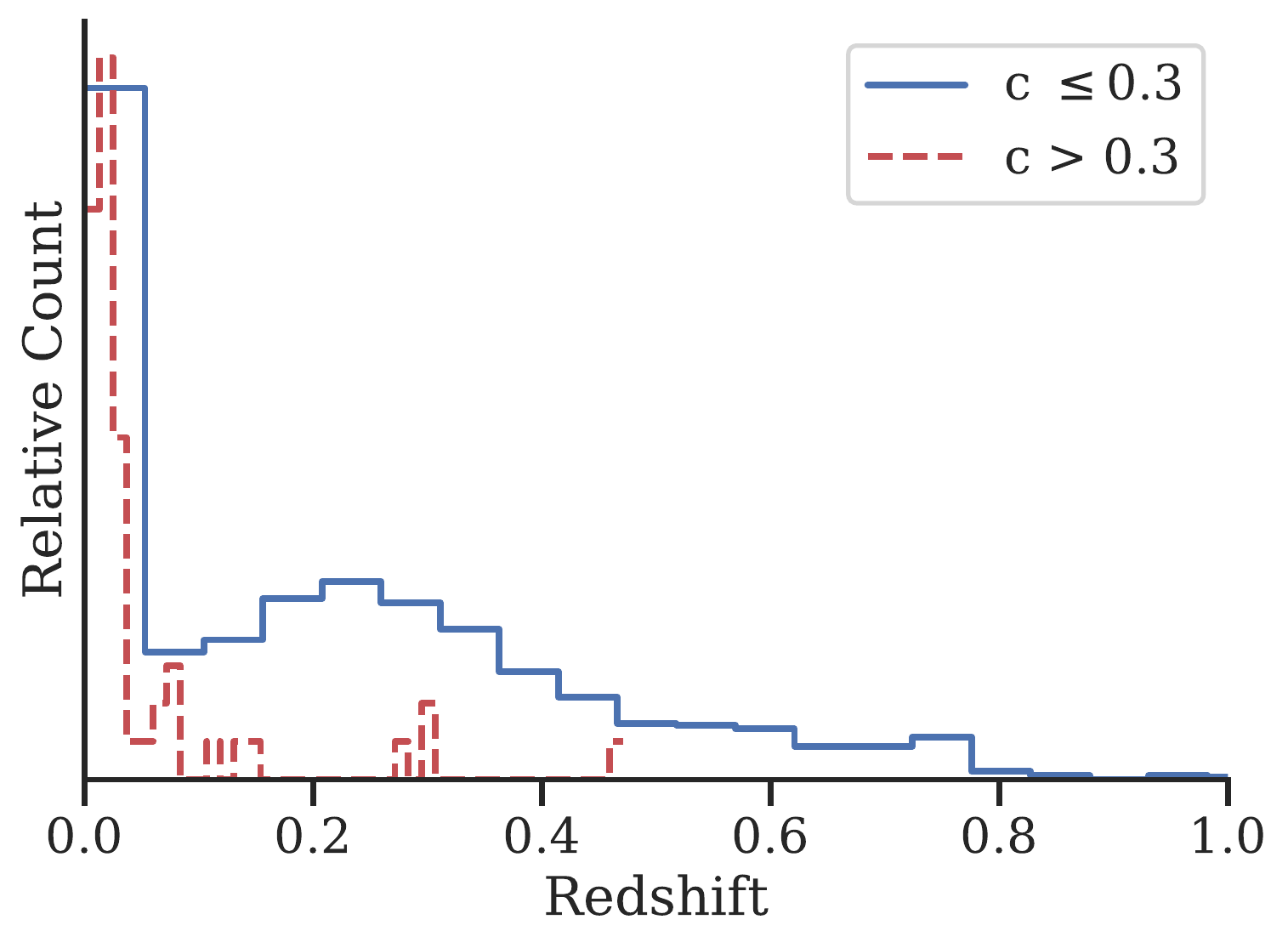}
     \end{subfigure}
    \caption{Distributions of \sn light-curve shape (\textit{left}), host galaxy stellar mass (\textit{center}), and redshift (\textit{right}) for the cosmological and highly reddened sub-samples.
    Using the two-sample Kolmogorov-Smirnov test, we see a difference in $x_1$ ($3.0\sigma$) and redshift ($>$5$\sigma$) between these two samples, but not in stellar mass ($0.5\sigma$).
    }
    \label{fig:red-vs-cosmo}
\end{figure*}

We show in \cref{fig:red-vs-cosmo} the light-curve shape ($x_1$), host galaxy stellar mass ($M_{\star}$, in units of $\log(M_{\sun})$), and redshift distributions for the cosmological sample and the highly reddened sample. The reddened sample's $x_1$ distribution changes significantly compared to the cosmological sample ($3.0\sigma$, via a two-sample Kolmogorov-Smirnov test), due to a relative increase of fast decliners ($x_1 < 0$). 
This shift in the $x_1$ population is not likely to be from an enhanced selection effect, since \sne with negative $x_1$ are fainter on average. We speculate that this shift may be the result of a change in the host-galaxy population, although not a different stellar mass distribution, as seen below. It has been previously shown that $x_1$ is correlated with host galaxy properties \citep[e.g.,][]{Hamuy2000}, and that for any given galaxy, the apparent $x_1$ distribution is smaller than it is for the \sne population as a whole \citep{Scolnic2020}. 
Alternatively, highly reddened \sne may have an enhanced SN1991bg-like population.
The redshift distributions also differ ($>$5$\sigma$). This is due to the increased effect of selection effects on highly reddened \sne, resulting in a redshift distribution that favours low redshifts. However, the $M_{\star}$ distributions are consistent between these two samples ($0.5\sigma$). 

\begin{figure}
    \centering
    \includegraphics[width=.9\columnwidth]{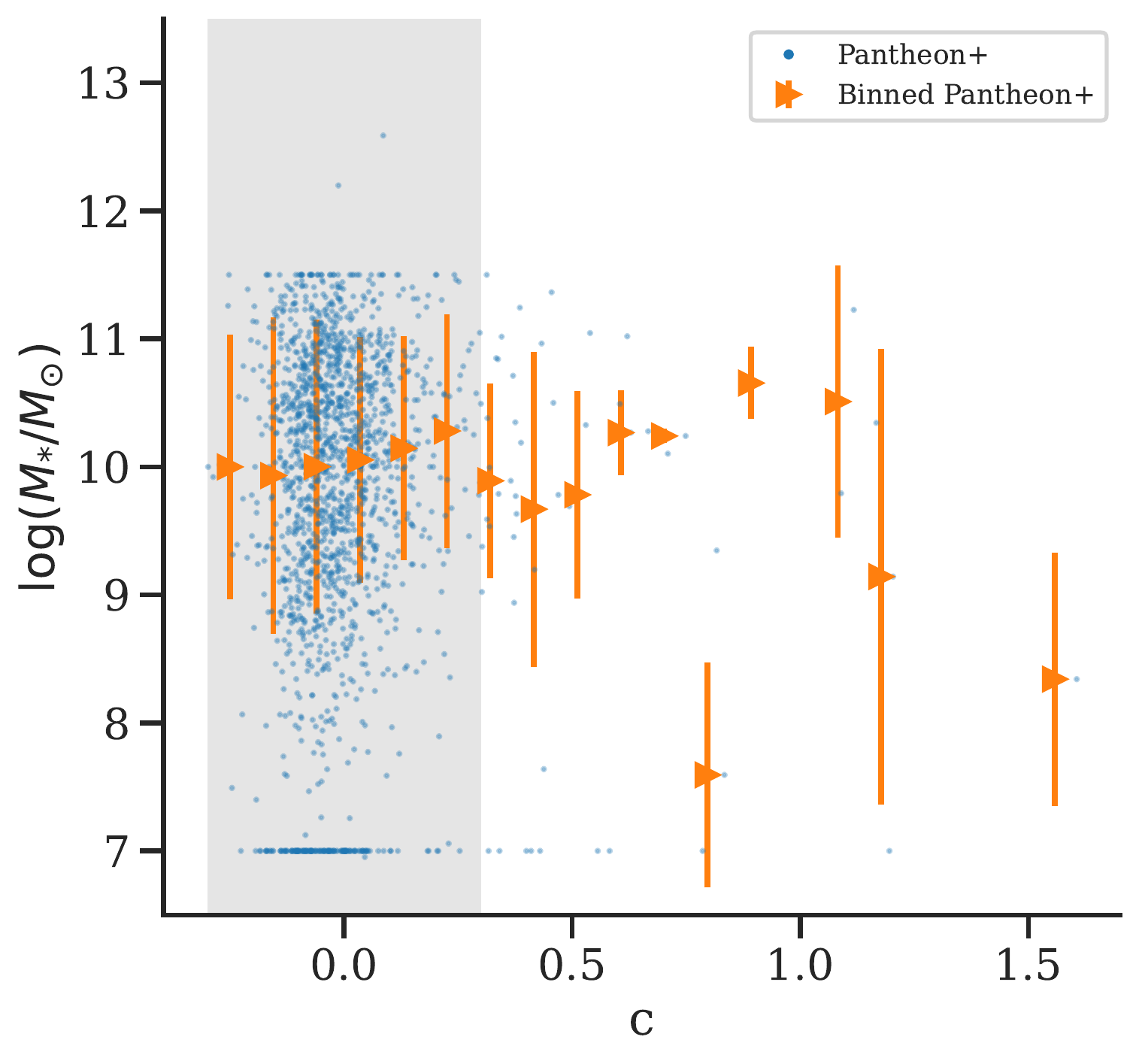}
    \caption{
    Host galaxy stellar mass as a function of \salt \co. 
    In bins of colour, the orange triangles show the median $\log(M_{*}/M_{\odot})$ and the uncertainty of the points show the robust scatter.
    At large values of \co, the relative decrease of \sne adds statistical noise, but the red \sne are not found in a substantially different host galaxy population than the cosmological sample (gray shaded region).
    The lines at 11.5 and 7 are artefacts of the stellar mass fitting process.
    A value of 7 was given when a mass was indeterminable due to faint photometry.
    A value of 11.5 was given when the mass is sufficiently large to be extremely unlikely to be less than 10. Some masses are larger than 11.5 because they are from surveys that don't follow the previous convention.
    }
    \label{fig:mass-colour}
\end{figure}

In \cref{fig:mass-colour}, we show the host galaxy stellar mass distribution as a function of \salt colour. At highly reddened colours, the relative decrease of \sne adds statistical noise, but the red \sne are not found in a substantially different host galaxy population than the cosmological sample.

\subsection{Simulations}

We use simulations of the \pplus data set generated by the SNANA simulation software \citep{Kessler2009, Kessler2019} with the \texttt{PIPPIN} \citep{Hinton2020} wrapper. A brief overview of SNANA is as follows: 1) fluxes are generated from a source model (\salt, \citealt{Taylor2021}), with noise and detection characterized in survey-specific methods alongside a rest-frame SED for each epoch that is subject to cosmological and galactic effects, 2) the SED is integrated for each filter to obtain broadband fluxes, which are further subjected to survey-specific measurement noise, and 3) candidate logic and spectroscopic identification efficiencies are applied. For this paper, we take our \pplus simulation inputs from \cite{Brout2022b}, a brief overview of which is presented in Table~2 of \cite{Popovic2022}.

\section{Method}\label{sec:method}

In this work, both \rv and $\beta$ represent a relationship between colour and luminosity.
$\beta$ is a fit parameter used with \sn data and empirical models like \salt, whereas \rv is a physical property and used in forward model simulations. 
Therefore, a colour dependence of \rv can be seen as a colour dependence of $\beta$.

We perform two types of fits on the \pplus data. The first model assumes a single $\beta$ for the whole sample. This is described in \cref{sec:method-single}. The second is a broken-linear model where $\beta$ can change after a certain colour value, to test the predictions of \citetalias{Mandel2011} and others that the colour-luminosity relationship is colour dependent (\cref{sec:method-broken}).
We also discuss how we measure the scatter around the colour-luminosity relation in \cref{sec:method-rms}. Finally, we present four models of the intrinsic scatter of \sne brightnesses, including possible colour dependence of \rv, we use in our forward model simulations (\cref{sec:method-scatters}). 

\subsection{Colour-luminosity Relationship} \label{sec:method-single}

We modify \cref{eqn:absmag-full} in order to define a non-colour corrected absolute magnitude (\absmag) per \sn:
\begin{equation}\label{eqn:absmag}
    \mathrm{M}' \equiv m_B - \mu(z, \mathrm{cosmology}) + \alpha x_1~.
\end{equation}

To focus this work on understanding $\beta$, we do not fit for either the cosmology or $\alpha$. Instead, we assume a flat \lcdm cosmology of $h=0.70$, $\Omega_M=0.3$, $\Omega_{\Lambda}=0.7$ and use $\alpha=0.15$. These are similar to those derived from the \pplus cosmological analysis \citep{Brout2022b}.  We note that in \cref{eqn:absmag} and throughout, no bias corrections are applied to any distance-like estimates.

\subsection{Modelling the Colour-Luminosity Relationship}\label{sec:method-broken}

We introduce two models of the colour-luminosity relation to fit $\mathrm{M}'$ defined in \cref{eqn:absmag}: a linear one and a broken-linear one.  Our linear one is defined such that
\begin{equation}\label{eqn:linear}
    \mathrm{M}'_{\mathrm{mod}}=\beta c~ + \mathrm{M}^{'}_0
\end{equation}
where $\mathrm{M}{'}_0$ is the absolute magnitude of a $c=0$ \sn and $\mathrm{M}'_{\mathrm{mod}}$ is the model's prediction for $\mathrm{M}'$.

For the broken-linear model, we build a \linmix-like \citep{Kelly2007} Bayesian Hierarchical model (\bhm) around the equation:
\begin{equation}\label{eqn:broken}
    \mathrm{M}'_{\mathrm{mod}}=\begin{cases}
          \beta c + \mathrm{M}^{'}_0 \hfill &\text{if} \, c \le c_{\mathrm{break}} \\
          (\beta + \Delta_{\beta})c + \mathrm{M}^{'}_0  - c_{\mathrm{break}}\Delta_{\beta} \quad &\text{if} \, c > c_{\mathrm{break}} \\
     \end{cases}
\end{equation}
where $\Delta_{\beta}$ is the change in the slope between the cosmological and highly reddened \sn.
The $- c_{\mathrm{break}}\Delta_{\beta}$ term allows for the piecewise function to remain continuous at $c=c_{\mathrm{break}}$.
In the fiducial analysis $c_{\mathrm{break}} = 0.3$.

This model is partially derived from \unity \citep{Rubin2015}. A full description of the \bhm used can be seen in Appendix \ref{sec:fullmodel}. To fit for $\beta$, we use a Python port\footnote{\url{https://github.com/jmeyers314/linmix}} of the IDL package \texttt{LINMIX\_ERR} \citep{Kelly2007}. 
We take advantage of two mathematical characteristics of \linmix. First, it is Bayesian and naturally provides uncertainties on the model parameters. Secondly, it handles two-dimensional measurement uncertainties. 

\subsection{Residual Scatter}\label{sec:method-rms}

The best fit results from the models presented in \cref{sec:method-single,sec:method-broken} will still result in residual scatter. The scatter in \sn absolute magnitude is a common diagnostic \citepalias[e.g.,][]{Brout2021}. For this work, we will calculate the root-mean-squared (RMS) scatter of the residuals as a function of colour to diagnose if this scatter is constant with colour, or if red \sne have an increased scatter. An increase in scatter as a function of colour is a prediction of the variable \rv model of \citetalias{Brout2021}.

\subsection{Forward Modelling Different Scatter Models}\label{sec:method-scatters}

Simulating \sn data sets require an empirical model to describe brightness variations among the \sn population, or intrinsic scatter. 
In this work, we use four different scatter models---\citealp{Guy2010} (\citetalias{Guy2010}), \citealp{Chotard2011} (\citetalias{Chotard2011}),
the \citealp{Popovic2022} (\citetalias{Popovic2022}) fits for \citetalias{Brout2021}, 
and \citealp{Mandel2011} (\citetalias{Mandel2011}). Both \citetalias{Guy2010} and \citetalias{Chotard2011} are spectral variation models, though they differ in the amount of variation ascribed to chromatic scatter. \citetalias{Guy2010} attributes approximately 70\% of scatter to achromatic effects, with the rest coming from chromatic sources. \citetalias{Chotard2011} only 25\% of scatter is achromatic, and the remaining 75\% is from chromatic effects.

In contrast to \citetalias{Guy2010} and \citetalias{Chotard2011}, \citetalias{Popovic2022} and \citetalias{Mandel2011} do not have an explicit SED variation; instead, \citetalias{Popovic2022} and \citetalias{Mandel2011} ascribe scatter to dust effects. 
We use the updated \citetalias{Brout2021} model parameters from \citetalias{Popovic2022} for this paper. This assumes no relationship between $E(B-V)$ and \rv. We implement the \citetalias{Mandel2011} model by replacing the $R_V$ and $E(B-V)$ relationships from \citetalias{Popovic2022} with those from \cite{Mandel2011}:
\begin{equation}
    \label{eqn:mandel}
    \frac{1}{\rv} = 0.36 + 0.14 A_V
\end{equation}
with a Gaussian scatter around this mean $\frac{1}{\rv}$ of $\sigma=0.04$ \citep{Mandel2011}.

\section{Results}\label{sec:results}

\begin{figure}
    \centering
    \includegraphics[width=.99\columnwidth]{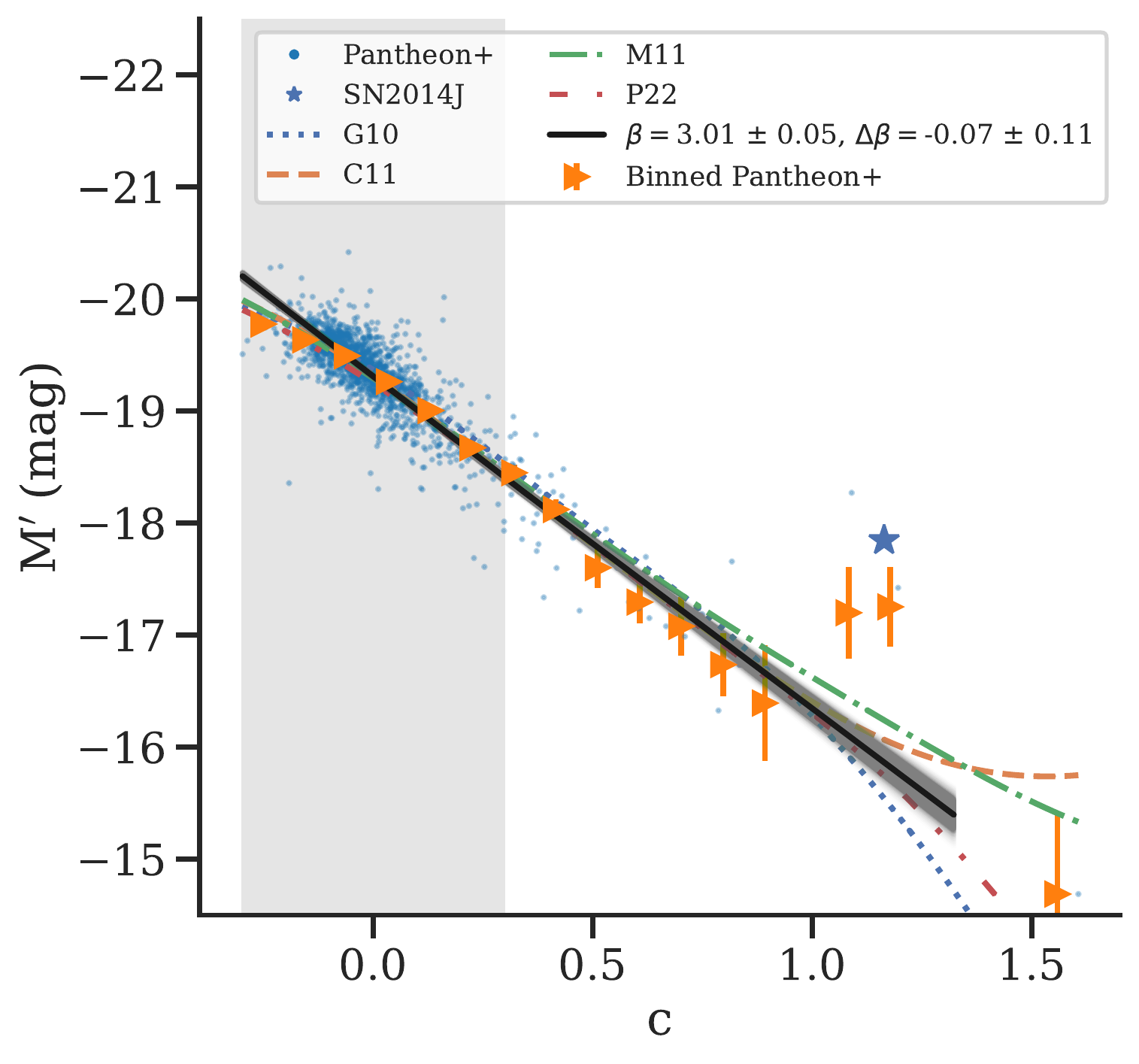}
    \caption{
      Colour-luminosity plot for our full sample (blue), including \Nmore highly reddened \sn.
      Bins in colour show the median $\mathrm{M}^{'}$ (with error bars showing the RMS divided by the square-root of the number of objects per bin, orange).
      The median (and $1\sigma$) broken-linear model is shown in black. We find $\beta = \rvcosmo$ and $\Delta\beta = \rvdelta$. This is consistent with only one slope across this whole range. 
      We also present the non-parametric kernel regression of the median luminosities for each scatter model (\citetalias{Guy2010}, blue-dotted line; \citetalias{Chotard2011}, orange-dashed line; \citetalias{Mandel2011}, green-dot-dashed line; \citetalias{Popovic2022}, red-dot-dot-dashed line.)
      We note that SN2014J (blue star) is significantly brighter than the trend, as expected for its low \rv value.
      The gray shaded region is the traditional cosmological sample.
      }
    \label{fig:colour-luminonsity-bs21}
\end{figure}

We show the colour-luminosity plot for the \pplus data set (including the highly reddened \sne) in \cref{fig:colour-luminonsity-bs21}. In addition to the data, we show equally sized colour bins of the median luminosity and root-mean-square (RMS) divided by the square-root of the number of objects per bin.
We also plot the non-parametric kernel regression of the median luminosities for each scatter model. 
We use a variant of Nadaraya-Watson kernel regression \citep{Nadaraya1964,Watson1964} that uses a local linear regression estimator, as implemented in the Python package \texttt{statsmodels} \citep{statsmodels2010}.

\subsection{Measuring $\beta$}

\begin{table}
    \centering
    \begin{tabular}{l|ccr}
        \hline \hline 
        $c_{\mathrm{break}}$ & $\beta$ & $\Delta_{\beta}$ & Significance \\
        \hline
        0.0 & $2.64 \pm 0.06$ & $0.49 \pm 0.08$ & 6.1$\sigma$\\
        0.2 & $2.96 \pm 0.05$ & $0.08 \pm 0.10$ & 0.8$\sigma$\\
        0.3 & $\rvcosmo$ & $\rvdelta$ & 0.6$\sigma$\\
        0.4 & $3.03 \pm 0.05$ & $-0.20 \pm 0.12$ & 1.7$\sigma$\\
        0.5 & $3.03 \pm 0.05$ & $-0.28 \pm 0.13$ & 2.2$\sigma$\\
        \hline
    \end{tabular}
    \caption{$\beta$ and $\Delta_{\beta}$ values for various broken-linear break points (Equation~\ref{eqn:broken}). The fiducial analysis breaks at $c=0.3$ because that is the typical cosmological cut.}
    \label{tab:betas}
\end{table}

Following the procedure described in \cref{sec:method-single},
we calculate a single $\beta$ with a median plus or minus the robust scatter: $\beta=\betaLINMIX$.
For the broken-linear \bhm, described in \cref{sec:method-broken}, we see a consistent slope above and below the $c=0.3$ split.
We find $\beta=\rvcosmo$ and $\Delta_{\beta}=\rvdelta$, resulting in a $\beta(c>0.3)=\rvred$. This is a $<$0.7$\sigma$ deviation from a constant linear trend.
The result of our \bhm{} can be seen plotted along with the data in \cref{fig:colour-luminonsity-bs21}.
All model parameters are well constrained, with $\beta$ and $\Delta_{\beta}$ being the only model parameters showing a notable correlation. For the full details of the fit, see Appendix \ref{sec:fitresults}.

We present the $\beta$ and $\Delta_{\beta}$ estimates for several values of $c_{\mathrm{break}}$ in \cref{tab:betas}. Between $0.2 \le c_{\mathrm{break}} \le 0.5$ there are no significant $\Delta_{\beta}$ values. However, there is a very significant shift in $\beta$ at $c=0$. A shift in $\beta$ at $c=0$ has been seen multiple times before \citep{Scolnic2014,Rubin2015}. This shift is thought to be a transition of the observed color being dominated from intrinsic color to dust.

\begin{figure}
    \centering
    \includegraphics[width=.99\columnwidth]{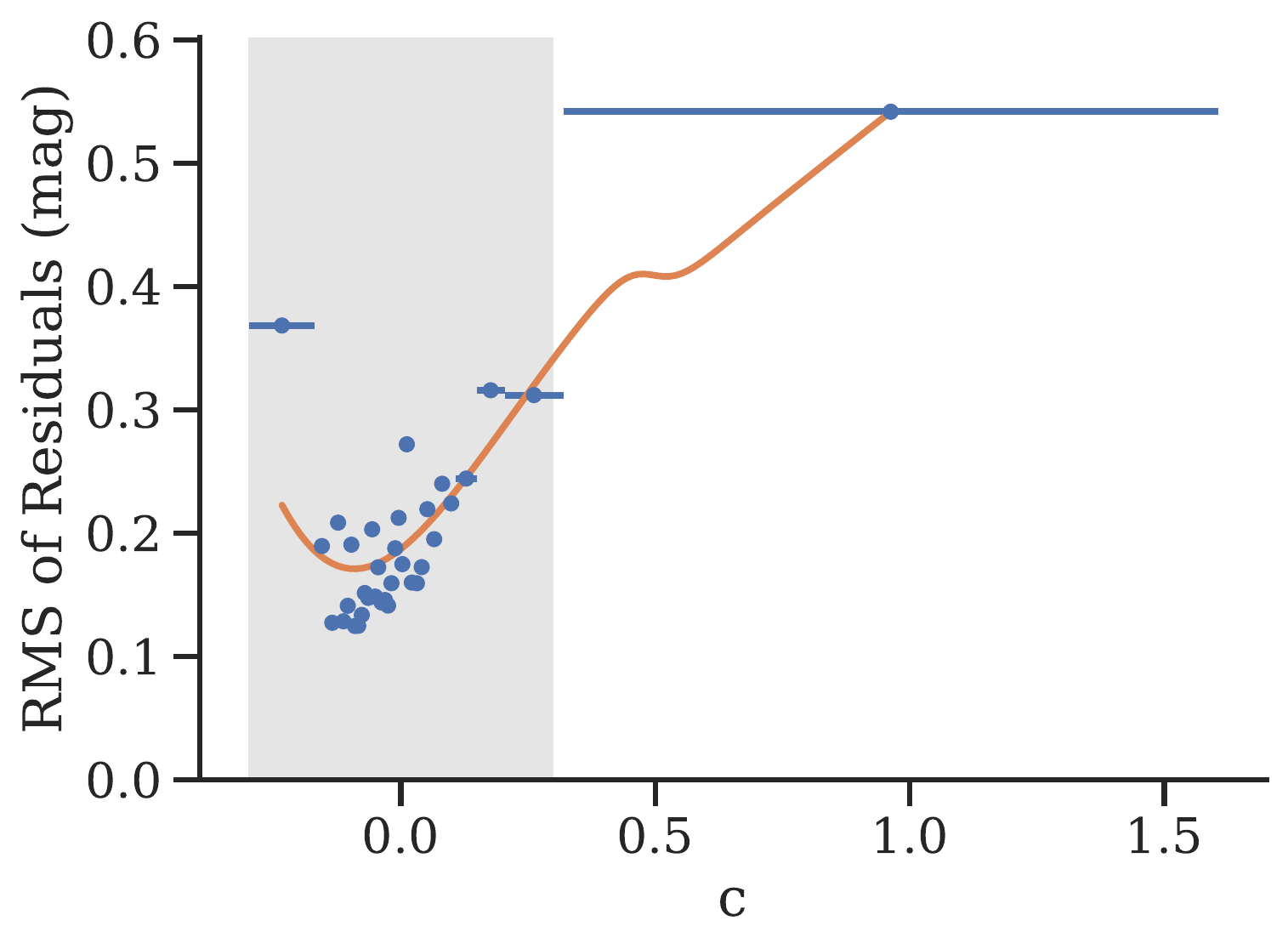}
    \caption{
      The root-mean-squared (RMS) scatter of the residuals between the data and the median broken linear model ($\beta=\rvcosmoMedian$ and $\beta(c>0.3)=\rvredMedian$).
      Each bin is evenly filled with 50 objects each.
      Highly reddened \sn have more scatter than the cosmological sample. However, they do appear to continue the trend seen in the red half ($c>0$) of the cosmological sample. 
      The full cosmological sample is denoted by the shaded region.
      A non-parametric kernel regression is plotted in orange. It smooths and interrelates between the points.
      We also calculated the robust scatter of the residual to see if this trend was dominated by outliers. Though the value of the scatter decreased, the trends seen in this plot did not change.
      }
    \label{fig:rms}
\end{figure}

In \cref{fig:rms}, we present the RMS scatter of the residuals vs colour. Each bin is evenly filled with 50 objects each.
This allows for robust measurements of the RMS while distinguishing the highly reddened ($c>0.3$) \sne from the cosmological sample.
Highly reddened \sn have larger scatter than the cosmological sample. However, they continue the trend seen in the red half ($c>0$) of the cosmological sample.
We use the same \texttt{statsmodels} kernel regression method to smooth and interpolate between the points.
This trend is robust to changes in bin size.
We also calculate the robust scatter of the residuals to verify that the observed trend is resilient to outliers. Though the value of the scatter decreases, the trends seen in this plot do not change.
An increasing RMS with colour is predicted by the \citetalias{Brout2021} model, but \citetalias{Guy2010} and \citetalias{Chotard2011} both assume a constant scale of remaining scatter with colour.

\subsection{Comparison of Forward Model Simulations}\label{sec:results-models}

We check the four scatter models described in \cref{sec:method-scatters} to determine which best matches the \pplus data. To quantify this agreement, we take inspiration from the metrics presented in \citetalias{Popovic2022}. We have one metric related to the distribution of $c$ and one related to the colour-luminosity relationship.

The first metric describes the relative number of \sne with red colours and compares the predictions from simulations using different scatter models to the data. By using the scatter models from \citetalias{Guy2010} and \citetalias{Chotard2011} with asymmetric Gaussian colour distributions \citep{Scolnic2016,Popovic2021}, we find that simulations with these models predict no $c>1$ \sn. This can be seen in colour histograms of the simulations in \cref{fig:colour}.
In fact, a $c>1$ \sn would be a $>$9$\sigma$ outlier. The \pplus data set has \Nextreamly \sne with $c>1$.  On the other hand, the \citetalias{Popovic2022} model is based on an exponential dust distribution, and therefore simulations with this model predict more \sne at redder colours, which better matches the data.

In order to quantify these colour-distribution predictions, we calculate a $\chi^2$ defined as:
\begin{equation}
    \chi^2_c = \sum_i^{N_{\mathrm{bins}}} \frac{(N_{\mathrm{data, i}} - N_{\mathrm{sim, i}})^2}{\sigma_{\mathrm{Poisson, i}}^2}
\end{equation}
where $N_i$ is the number of observed \sn per bin in either the data or simulation, and $\sigma_{\mathrm{Poisson}}$ is the typical Poisson counting uncertainty on the data bins ($\sqrt{N_{\mathrm{data}}}$).

\begin{table}
    \centering
    \begin{tabular}{l|ccc}
        \hline \hline 
         & 
        $\chi^2_c$ & $\chi^2_{\mathrm{M}^{'}}$ & $\chi^2_{\mathrm{M}^{'}},~c < 1.0$ \\
         & ($N_{\mathrm{bins}}=20)$ & ($N=\Ntotal$) & ($N=1680$)\\
        \hline
        \citetalias{Popovic2022} & 38.3 & 1566.8 & 1537.9\\
        \citetalias{Mandel2011} & 58.6  & 1568.8 & 1551.2\\
        \citetalias{Chotard2011} & 45.2 & 1561.6 & 1538.4\\
        \citetalias{Guy2010} &  45.2 & 1580.5 & 1537.6\\
        \hline
    \end{tabular}
    \caption{$\chi^2$ Between Forward Model Simulations and the \pplus Data Set}
    \label{tab:chi}
\end{table}

We also calculate a $\chi^2$ for the colour-luminosity relationship. This compares the data ($\mathrm{M}^{'}_i$) to the kernel regression smoothed luminosity ($\mathrm{M}^{'}_{\mathrm{sim}}(c_i)$) of each simulation with the uncertainty coming from the measured scatter in the data: 
\begin{equation}
    \chi^2_{\mathrm{M}^{'}} = \sum^{N}_i\frac{\left[\mathrm{M}^{'}_i - \mathrm{M}^{'}_{\mathrm{sim}}(c_i) - \Delta \mathrm{M}_0\right]^2}
    {\mathrm{RMS}(c_i)}
\end{equation}
where $\mathrm{RMS}(c_i)$ is the kernel regression smoothed for a given \sn colour ($c_i$), as presented in \cref{fig:rms}.
$\Delta \mathrm{M}_0$ is a single number for the whole dataset and is a shift in absolute magnitude of an $x_1=c=0$ \sn. This is calculated as the median of $(\mathrm{M}^{'}_i - \mathrm{M}^{'}_{\mathrm{sim}}(c_i))$ and is up to 0.1~mag, due to the different definitions used when defining different scatter models.

Our results are shown in \cref{tab:chi}. When looking at the colour distribution, \citetalias{Popovic2022} matches the data the best. \citetalias{Guy2010} and \citetalias{Chotard2011} match the data better than \citetalias{Mandel2011}, even though they see $c>1$ \sn as $>$9$\sigma$ outliers. For the colour-luminosity relationship, \citetalias{Chotard2011} does the best, but \citetalias{Popovic2022} and \citetalias{Mandel2011} have a similar $\chi^2$. Cutting the low statistic $c>1.0$ data range, \citetalias{Guy2010}, \citetalias{Popovic2022} and \citetalias{Chotard2011} are all within a $\Delta \chi^2$ of $<$1.

\section{Discussion}\label{sec:discussion}

\subsection{Low \rv \sn in the Literature}

Though we find a consistent $\beta$ for \pplus across a wide range of \sn colours, there are well measured highly reddened individual \sn, such as SN2014J, with low \rv values \citep[$\rv \sim 1.5$,][]{Amanullah2015}.
In \cref{fig:colour-luminonsity-bs21}, we show that SN2014J is above the measured $\beta$.
We see individual \sne with low \rv, even though the mean for the data set ($\beta$) is higher, which implies there is a range of \rv values.

\begin{table}
    \centering
    \begin{tabular}{ l | ccc | cr }
        \hline\hline
        \sn & c & Stellar Mass & $\Delta \mathrm{mag}$ & \rv & Reference\\
         &  & $\log(M_*/M_{\sun})$& mag &  & \\
        \hline
        1999cl\tablenotemark{\footnotesize{*}} & 1.09 & 9.9 & -1.25 & $1.55 \pm 0.08$ & K06\\
        2002bo\tablenotemark{\footnotesize{*}} & 0.32 & \nodata & 0.77 & $1.00 \pm 0.10$ & ER06\\
            & & & & $1.22^{+0.26}_{-0.21}$ & P13\\
            & & & & $1.1$ & C16\\
        2006X\tablenotemark{\footnotesize{*}} & 1.17 & 10.3 &  -0.46  & $1.31^{+0.08}_{-0.1}$ & P13\\ 
        2008fp\tablenotemark{\footnotesize{*}} & 0.30 & 10.5 & 0.77 & $1.20^{+0.26}_{-0.14}$  & P13\\
        2012cg & 0.11 & 9.6 & 1.60 & $2.7^{+0.9}_{-0.7}$ & A15\\
        2014J\tablenotemark{\footnotesize{*}} & 1.16 & \nodata & -0.71 & $1.4 \pm 0.1$ & A15\\
        \hline
    \end{tabular}
    \caption{\pplus \sne with \rv Measurements in the Literature.
    Stellar mass and $c$ are from the \pplus data release. $\Delta \mathrm{mag}$ is the residual between M$'$ and the trend line presented in \cref{fig:colour-luminonsity-bs21}.\\
    \textit{References:} A15 \citep{Amanullah2015}; C16 \citep{Cikota2016}; ER06 \citep{Elias-Rosa2006}; K06 \citep{Krisciunas2006};  P13 \citep{Phillips2013}.\\
    $^{*}$A \sn with $z<0.01$ and $c>0.3$.
    }
    \label{tab:rvs}
\end{table}

In \cref{tab:rvs}, we present a table of six \sne with \rv measurements from the literature. In this table we include the colour and host galaxy stellar mass from \pplus as well as the residual from the median fit presented in \cref{fig:colour-luminonsity-bs21} ($\beta=\rvcosmo$ and $\Delta_{\beta}=\rvdelta$).
SN1999cl, SN2006X, SN2012cg, and SN2014J all follow the expected trend of brighter residuals for lower \rv.
However, a conclusion is hard to draw, both due to the low number of objects and that SN2002bo and SN2008fp do not follow this trend.

\subsection{Improving \h}

Previous studies have raised that a variable \rv may affect measurements of \h.
\citet{Mortsell2021} derived a colour-luminosity relation for each Cepheid host galaxy individually, though with large uncertainties.
Directly measuring the line-of-sight \rv for Cepheid and \sne, especially with dust dominated, highly reddened \sne, would provide an interesting comparison and illuminate whether the dust around \sne is interstellar or circumstellar.

\begin{table}
    \centering
    \begin{tabular}{lccr}
        \hline\hline
        \sn & Host Galaxy & Redshift & c \\
        \hline
        2014J   & M82 & 0.00138 & 1.16 \\
        1996ai & NGC 5005 & 0.00451 & 1.61\\
        2002bo  & NGC 3190 & 0.00613 & 0.32 \\
        2017drh & NGC 6384 & 0.00616 & 1.22 \\
        2008fp  & ESO 428-G14 & 0.00618 & 0.31 \\
        2006X & M100 & 0.00664 & 1.17 \\
        1997dt  & NGC 7448 & 0.00683 & 0.47 \\
        2007bm  & NGC 3672 & 0.00727 & 0.38 \\
        1996bk  & NGC 5308 & 0.00779 & 0.33 \\
        1999cl  & M88 & 0.00850 & 1.09 \\
        \hline
    \end{tabular}
    \caption{Possible \h Calibrators. We present all $c > 0.3$ \sne with $z < 0.01$ (N=\NnewCalibrators).
    A total of \NnewCalibratorsAll highly-reddened \sne are available if calibration is possible to $z < 0.015$ or $\mu \lesssim 34 \un{mag}$. This table excludes \pplus \sne with $c<0.3$ but are not typical \h calibrators, such as SN1989B ($c=0.298 \pm 0.038$) and SN1998bu ($c=0.261 \pm 0.025$).}
    \label{tab:calibrators}
\end{table}

Furthermore, without the $c<0.3$ cut, there is an increased number of \h calibrators ($z<0.01$ \sn where either Cepheid or Tip of the Red Giant Branch distances are possible, \citealt{Freedman2019}) by \NnewCalibrators---see \cref{tab:calibrators} for a full list. There are \NnewCalibratorsAll possible calibrators if we extend this to $z<0.015$ or $\mu \sim 34$.
The latest \h measurement, by the SH0ES collaboration, contains 42 Cepheid-calibrated \sn \citep{Riess2022}. The rate of new \sn calibrators is $\sim 1/\mathrm{year}$, so these highly reddened \sn pose a promising method to increase the precision of the local measurement of \h, which can be forecasted using the precision per \sn with redder colours as seen in \cref{fig:rms}.

\section{Conclusions}\label{sec:conclusions}

Though \salt was optimized in the cosmological colour range ($-0.3 < c < 0.3$) it still has good agreement with the data out to $c > 1$.
We see no evidence for future research to cut at $c = 0.3$, but suggest consideration of the use of \sn with $c \lesssim 1$.

We find a consistent slope between $c<0.3$ and $c>0.3$ ($\Delta_{\beta}=\rvdelta$).
Measurements of redder \sne do have a larger post standardization scatter, but they follow a trend between RMS and colour also seen in the cosmological sample and explained by \citetalias{Brout2021}.
A limit in colour is still needed to avoid low number statistics where we will not be able to constrain the variability in \rv. Highly reddened \sn can be included in cosmological samples, or at least used for constraining systematic uncertainties in the more statistically powerful samples.

Our results strongly support that $R_V$ varies across the \sne population, and must be accounted for in cosmological analyses.
The \citetalias{Popovic2022} scatter model matches the colour distribution best, with \citetalias{Chotard2011}, \citetalias{Popovic2022} and \citetalias{Mandel2011} matching the colour-luminosity relationship about equally.
Additionally, the correlation between the spectroscopicly measured \rv of \sne and their residual scatter is inconclusive. However, we note this is from only six \sne and would benefit from a larger sample.

Finally, we want to conclude by reiterating that highly reddened \sne are highly affected by selection effects. A good model of these effects is required for any study of them or perform analyses with them.

\section*{Acknowledgements}
The authors would like to thank 
David Rubin for discussions on building Bayesian Hierarchical models and comments on an early draft. 
The authors would like to thank the anonymous referee for their time and attention. Their comments and suggestions improved the clarity and value of this paper.
This work was completed, in part, with resources provided by the University of Chicago’s Research Computing Center.
B.M.R. and D.S. are supported, in part, by the National Aeronautics and Space Administration (NASA) under Contract No.~NNG17PX03C, issued through the Roman Science Investigation Teams Programme.
D.S. is supported by DOE grant DE-SC0010007, DE-SC0021962 and the David and Lucile Packard Foundation. 
D.S. and D.B. thank the John Templeton Foundation.
D.B. acknowledges support for this work provided by NASA through NASA Hubble Fellowship grant HST-HF2-51430.001 awarded by the Space Telescope Science Institute (STScI), which is operated by the Association of Universities for Research in Astronomy, Inc., for NASA, under contract NAS5-26555.

The research presented in this paper used the following software packages: 
ArviZ \citep{arviz_2019},
Astropy \citep{Astropy2013,Astropy2018}, 
corner.py \citep{Foreman-Mackey2016}, 
Graphviz (\url{https://www.graphviz.org}),
\textsc{LINMIX} \citep{Kelly2007},
Matplotlib \citep{matplotlib}, 
Numpy \citep{numpy}, 
the Open Supernova Catalog \citep{OpenSNCatalog},
PyMC3 \citep{pymc3},
Pandas \citep{pandas}, 
PIPPIN \citep{Hinton2020},
The Plotter,
Python, 
SciPy \citep{scipy}, 
Seaborn \citep{seaborn},
SNANA \citep{Kessler2009a,Kessler2017},
statsmodels \citep{statsmodels2010}, 
and
Theano \citep{theano2016}.

\section*{Data Availability}

This work uses the \pplus data set. The \pplus data set is available at \href{https://pantheonplussh0es.github.io}{\url{pantheonplussh0es.github.io}}. The associated PyMC3 model can be found at \href{https://github.com/benjaminrose/Investigating-Red-SN}{\url{github.com/benjaminrose/Investigating-Red-SN}}. Any other data requests can be directed to the lead author.

\bibliographystyle{mnras}
\bibliography{library}

\providecommand{\noopsort}[1]{}
\begin{thebibliography}{}
\makeatletter
\relax
\def\mn@urlcharsother{\let\do\@makeother \do\$\do\&\do\#\do\^\do\_\do\%\do\~}
\def\mn@doi{\begingroup\mn@urlcharsother \@ifnextchar [ {\mn@doi@}
  {\mn@doi@[]}}
\def\mn@doi@[#1]#2{\def\@tempa{#1}\ifx\@tempa\@empty \href
  {http://dx.doi.org/#2} {doi:#2}\else \href {http://dx.doi.org/#2} {#1}\fi
  \endgroup}
\def\mn@eprint#1#2{\mn@eprint@#1:#2::\@nil}
\def\mn@eprint@arXiv#1{\href {http://arxiv.org/abs/#1} {{\tt arXiv:#1}}}
\def\mn@eprint@dblp#1{\href {http://dblp.uni-trier.de/rec/bibtex/#1.xml}
  {dblp:#1}}
\def\mn@eprint@#1:#2:#3:#4\@nil{\def\@tempa {#1}\def\@tempb {#2}\def\@tempc
  {#3}\ifx \@tempc \@empty \let \@tempc \@tempb \let \@tempb \@tempa \fi \ifx
  \@tempb \@empty \def\@tempb {arXiv}\fi \@ifundefined
  {mn@eprint@\@tempb}{\@tempb:\@tempc}{\expandafter \expandafter \csname
  mn@eprint@\@tempb\endcsname \expandafter{\@tempc}}}

\bibitem[\protect\citeauthoryear{{Al-Rfou} et~al.,}{{Al-Rfou}
  et~al.}{2016}]{theano2016}
{Al-Rfou} R.,  et~al., 2016, arXiv e-prints, abs/1605.02688

\bibitem[\protect\citeauthoryear{Amanullah et~al.,}{Amanullah
  et~al.}{2015}]{Amanullah2015}
Amanullah R.,  et~al., 2015, \mn@doi [Mon. Not. R. Astron. Soc.]
  {10.1093/mnras/stv1505}, 453, 3301

\bibitem[\protect\citeauthoryear{{Astropy Collaboration}}{{Astropy
  Collaboration}}{2013}]{Astropy2013}
{Astropy Collaboration} 2013, A\&A, 558, A33

\bibitem[\protect\citeauthoryear{{Astropy Collaboration}}{{Astropy
  Collaboration}}{2018}]{Astropy2018}
{Astropy Collaboration} 2018, \mn@doi [Astronomical Journal]
  {10.3847/1538-3881/aabc4f}, \href
  {https://ui.adsabs.harvard.edu/abs/2018AJ....156..123A} {156, 123}

\bibitem[\protect\citeauthoryear{Betoule et~al.,}{Betoule
  et~al.}{2014}]{Betoule2014}
Betoule M.,  et~al., 2014, A\&A, 568, A22

\bibitem[\protect\citeauthoryear{Brout \& Scolnic}{Brout \&
  Scolnic}{2021}]{Brout2021}
Brout D.,  Scolnic D.,  2021, ApJ, 909, 26

\bibitem[\protect\citeauthoryear{Brout et~al.,}{Brout
  et~al.}{2021}]{Brout2022a}
Brout D.,  et~al., 2021, arXiv:2112.03864 [astro-ph]

\bibitem[\protect\citeauthoryear{Brout, Scolnic, Vincenzi, Dwomoh, Lidman,
  Riess, Ali  \& Dai}{Brout et~al.}{2022}]{Brout2022b}
Brout D.,  Scolnic D.,  Vincenzi M.,  Dwomoh A.,  Lidman C.,  Riess A.,  Ali
  N.,   Dai M.,  2022, in prep.

\bibitem[\protect\citeauthoryear{Brownsberger, Brout, Scolnic, Stubbs  \&
  Riess}{Brownsberger et~al.}{2021}]{Brownsberger2021}
Brownsberger S.,  Brout D.,  Scolnic D.,  Stubbs C.~W.,   Riess A.~G.,  2021,
  arXiv e-prints, \href {https://ui.adsabs.harvard.edu/abs/2021arXiv211003486B}
  {p. arXiv:2110.03486}

\bibitem[\protect\citeauthoryear{Burns et~al.,}{Burns et~al.}{2014}]{Burns2014}
Burns C.~R.,  et~al., 2014, \mn@doi [ApJ] {10.1088/0004-637X/789/1/32}, 789, 32

\bibitem[\protect\citeauthoryear{Cardelli, Clayton  \& Mathis}{Cardelli
  et~al.}{1989}]{Cardelli1989a}
Cardelli J.~A.,  Clayton G.~C.,   Mathis J.~S.,  1989, \mn@doi [The
  Astrophysical Journal] {10.1086/167900}, 345, 245

\bibitem[\protect\citeauthoryear{Carr, Davis, Scolnic, Said, Brout, Peterson
  \& Kessler}{Carr et~al.}{2021}]{Carr2021}
Carr A.,  Davis T.~M.,  Scolnic D.,  Said K.,  Brout D.,  Peterson E.~R.,
  Kessler R.,  2021, arXiv e-prints, \href
  {https://ui.adsabs.harvard.edu/abs/2021arXiv211201471C} {p. arXiv:2112.01471}

\bibitem[\protect\citeauthoryear{Chen et~al.,}{Chen et~al.}{2022}]{Chen2022}
Chen R.,  et~al., 2022, Measuring {{Cosmological Parameters}} with {{Type Ia
  Supernovae}} in {{redMaGiC}} Galaxies (\mn@eprint {arXiv} {2202.10480})

\bibitem[\protect\citeauthoryear{Chotard et~al.,}{Chotard
  et~al.}{2011}]{Chotard2011}
Chotard N.,  et~al., 2011, \mn@doi [A\&A] {10.1051/0004-6361/201116723}, 529,
  L4

\bibitem[\protect\citeauthoryear{Cikota, Deustua  \& Marleau}{Cikota
  et~al.}{2016}]{Cikota2016}
Cikota A.,  Deustua S.,   Marleau F.,  2016, \mn@doi [Astrophysical Journal]
  {10.3847/0004-637X/819/2/152}, \href
  {https://ui.adsabs.harvard.edu/abs/2016ApJ...819..152C} {819, 152}

\bibitem[\protect\citeauthoryear{{DES Collaboration} et~al.,}{{DES
  Collaboration} et~al.}{2019}]{DESCollaboration2019}
{DES Collaboration} et~al., 2019, \mn@doi [ApJL] {10.3847/2041-8213/ab04fa},
  872, L30

\bibitem[\protect\citeauthoryear{De~Marchi, Panagia  \& Milone}{De~Marchi
  et~al.}{2021}]{DeMarchi2021}
De~Marchi G.,  Panagia N.,   Milone A.~P.,  2021, arXiv e-prints, \href
  {https://ui.adsabs.harvard.edu/abs/2021arXiv210913914D} {p. arXiv:2109.13914}

\bibitem[\protect\citeauthoryear{{Elias-Rosa} et~al.,}{{Elias-Rosa}
  et~al.}{2006}]{Elias-Rosa2006}
{Elias-Rosa} N.,  et~al., 2006, \mn@doi [Monthly Notices of the Royal
  Astronomical Society] {10.1111/j.1365-2966.2006.10430.x}, \href
  {https://ui.adsabs.harvard.edu/abs/2006MNRAS.369.1880E} {369, 1880}

\bibitem[\protect\citeauthoryear{{Foreman-Mackey}}{{Foreman-Mackey}}{2016}]{Foreman-Mackey2016}
{Foreman-Mackey} D.,  2016, \mn@doi [JOSS] {10.21105/joss.00024}, 24

\bibitem[\protect\citeauthoryear{Freedman et~al.,}{Freedman
  et~al.}{2019}]{Freedman2019}
Freedman W.~L.,  et~al., 2019, \mn@doi [ApJ] {10.3847/1538-4357/ab2f73}, 882,
  34

\bibitem[\protect\citeauthoryear{Garnavich et~al.,}{Garnavich
  et~al.}{1998}]{Garnavich1998b}
Garnavich P.~M.,  et~al., 1998, ApJ, 509, 74

\bibitem[\protect\citeauthoryear{{Gonzalez-Gaitan}, {\noopsort{jaeger}}{de
  Jaeger}, Galbany, Mourao, {Paulina-Afonso}  \& Filippenko}{{Gonzalez-Gaitan}
  et~al.}{2021}]{Gonzalez-Gaitan2021}
{Gonzalez-Gaitan} S.,  {\noopsort{jaeger}}{de Jaeger} T.,  Galbany L.,  Mourao
  A.,  {Paulina-Afonso} A.,   Filippenko A.~V.,  2021, \mn@doi [Monthly Notices
  of the Royal Astronomical Society] {10.1093/mnras/stab2802}, 508, 4656

\bibitem[\protect\citeauthoryear{Goobar}{Goobar}{2008}]{Goobar2008}
Goobar A.,  2008, \mn@doi [The Astrophysical Journal Letters] {10.1086/593060},
  \href {https://ui.adsabs.harvard.edu/abs/2008ApJ...686L.103G} {686, L103}

\bibitem[\protect\citeauthoryear{Guillochon, Parrent, Kelley  \&
  Margutti}{Guillochon et~al.}{2017}]{OpenSNCatalog}
Guillochon J.,  Parrent J.,  Kelley L.~Z.,   Margutti R.,  2017, \mn@doi
  [Astrophysical Journal] {10.3847/1538-4357/835/1/64}, \href
  {https://ui.adsabs.harvard.edu/abs/2017ApJ...835...64G} {835, 64}

\bibitem[\protect\citeauthoryear{Gull}{Gull}{1989}]{Gull1989}
Gull S.~F.,  1989, in An {{International Book Series}} on {{The Fundamental
  Theories}} of {{Physics}}: {{Their Clarification}}, {{Development}} and
  {{Application}}, Vol.~36, Maximum {{Entropy}} and {{Bayesian Methods}}.
  {{Fundamental Theories}} of {{Physics}}.
{Springer, Dordrecht}

\bibitem[\protect\citeauthoryear{Guy et~al.,}{Guy et~al.}{2007}]{Guy2007}
Guy J.,  et~al., 2007, \mn@doi [A\&A] {10.1051/0004-6361:20066930}, 466, 11

\bibitem[\protect\citeauthoryear{Guy et~al.,}{Guy et~al.}{2010}]{Guy2010}
Guy J.,  et~al., 2010, \mn@doi [A\&A] {10.1051/0004-6361/201014468}, 523, A7

\bibitem[\protect\citeauthoryear{Hamuy, Trager, Pinto, Phillips, Schommer,
  Ivanov  \& Suntzeff}{Hamuy et~al.}{2000}]{Hamuy2000}
Hamuy M.,  Trager S.~C.,  Pinto P.~A.,  Phillips M.~M.,  Schommer R.~A.,
  Ivanov V.,   Suntzeff N.~B.,  2000, \mn@doi [AJ] {10.1086/301527}, 120, 1479

\bibitem[\protect\citeauthoryear{Harris et~al.,}{Harris et~al.}{2020}]{numpy}
Harris C.~R.,  et~al., 2020, \mn@doi [Nature] {10.1038/s41586-020-2649-2}, 585,
  357

\bibitem[\protect\citeauthoryear{Hinton \& Brout}{Hinton \&
  Brout}{2020}]{Hinton2020}
Hinton S.,  Brout D.,  2020, \mn@doi [JOSS] {10.21105/joss.02122}, 5, 2122

\bibitem[\protect\citeauthoryear{Huang et~al.,}{Huang et~al.}{2017}]{Huang2017}
Huang X.,  et~al., 2017, \mn@doi [The Astrophysical Journal]
  {10.3847/1538-4357/836/2/157}, \href
  {https://ui.adsabs.harvard.edu/abs/2017ApJ...836..157H} {836, 157}

\bibitem[\protect\citeauthoryear{Hunter}{Hunter}{2007}]{matplotlib}
Hunter J.~D.,  2007, \mn@doi [CSE] {10.1109/MCSE.2007.55}, 9, 90

\bibitem[\protect\citeauthoryear{Jeffreys}{Jeffreys}{1946}]{Jeffreys1946}
Jeffreys H.,  1946, \mn@doi [Proceedings of the Royal Society of London Series
  A] {10.1098/rspa.1946.0056}, \href
  {https://ui.adsabs.harvard.edu/abs/1946RSPSA.186..453J} {186, 453}

\bibitem[\protect\citeauthoryear{Jha, Riess  \& Kirshner}{Jha
  et~al.}{2007}]{Jha2007}
Jha S.,  Riess A.~G.,   Kirshner R.~P.,  2007, \mn@doi [ApJ] {10.1086/512054},
  659, 122

\bibitem[\protect\citeauthoryear{Jones, Oliphant, Peterson  et~al.}{Jones
  et~al.}{2001}]{scipy}
Jones E.,  Oliphant T.,  Peterson P.,   et~al., 2001, Nature Methods,
  arXiv:1907.10121

\bibitem[\protect\citeauthoryear{Kelly}{Kelly}{2007}]{Kelly2007}
Kelly B.~C.,  2007, \mn@doi [ApJ] {10.1086/519947}, \href
  {https://ui.adsabs.harvard.edu/abs/2007ApJ...665.1489K} {665, 1489}

\bibitem[\protect\citeauthoryear{Kessler \& Scolnic}{Kessler \&
  Scolnic}{2017}]{Kessler2017}
Kessler R.,  Scolnic D.,  2017, \mn@doi [ApJ] {10.3847/1538-4357/836/1/56},
  836, 56

\bibitem[\protect\citeauthoryear{Kessler et~al.,}{Kessler
  et~al.}{2009a}]{Kessler2009a}
Kessler R.,  et~al., 2009a, \mn@doi [PASP] {10.1086/605984}, 121, 1028

\bibitem[\protect\citeauthoryear{Kessler et~al.,}{Kessler
  et~al.}{2009b}]{Kessler2009}
Kessler R.,  et~al., 2009b, \mn@doi [The Astrophysical Journal Supplement
  Series] {10.1088/0067-0049/185/1/32}, 185, 32

\bibitem[\protect\citeauthoryear{Kessler et~al.,}{Kessler
  et~al.}{2019}]{Kessler2019}
Kessler R.,  et~al., 2019, \mn@doi [MNRAS] {10.1093/mnras/stz463}, \href
  {https://ui.adsabs.harvard.edu/abs/2019MNRAS.485.1171K} {485, 1171}

\bibitem[\protect\citeauthoryear{Krisciunas, Prieto, Garnavich, Riley, Rest,
  Stubbs  \& McMillan}{Krisciunas et~al.}{2006}]{Krisciunas2006}
Krisciunas K.,  Prieto J.~L.,  Garnavich P.~M.,  Riley J.-L.~G.,  Rest A.,
  Stubbs C.,   McMillan R.,  2006, \mn@doi [The Astronomical Journal]
  {10.1086/499523}, \href
  {https://ui.adsabs.harvard.edu/abs/2006AJ....131.1639K} {131, 1639}

\bibitem[\protect\citeauthoryear{Kumar, Carroll, Hartikainen  \& Martin}{Kumar
  et~al.}{2019}]{arviz_2019}
Kumar R.,  Carroll C.,  Hartikainen A.,   Martin O.~A.,  2019, \mn@doi [The
  Journal of Open Source Software] {10.21105/joss.01143}

\bibitem[\protect\citeauthoryear{Mandel, Narayan  \& Kirshner}{Mandel
  et~al.}{2011}]{Mandel2011}
Mandel K.~S.,  Narayan G.,   Kirshner R.~P.,  2011, \mn@doi [ApJ]
  {10.1088/0004-637X/731/2/120}, 731, 120

\bibitem[\protect\citeauthoryear{McKinney}{McKinney}{2010}]{pandas}
McKinney W.,  2010, Data {{Structures}} for {{Statistical Computing}} in
  {{Python}}

\bibitem[\protect\citeauthoryear{Meldorf et~al.,}{Meldorf
  et~al.}{2022}]{Meldorf2022}
Meldorf C.,  et~al., 2022, The {{Dark Energy Survey Supernova Program}}
  Results: {{Type Ia Supernova}} Brightness Correlates with Host Galaxy Dust
  (\mn@eprint {arXiv} {2206.06928})

\bibitem[\protect\citeauthoryear{Mortsell, Goobar, Johansson  \&
  Dhawan}{Mortsell et~al.}{2021}]{Mortsell2021}
Mortsell E.,  Goobar A.,  Johansson J.,   Dhawan S.,  2021, arXiv e-prints,
  \href {https://ui.adsabs.harvard.edu/abs/2021arXiv210511461M} {p.
  arXiv:2105.11461}

\bibitem[\protect\citeauthoryear{Mosher et~al.,}{Mosher
  et~al.}{2014}]{Mosher2014}
Mosher J.,  et~al., 2014, \mn@doi [ApJ] {10.1088/0004-637X/793/1/16}, 793, 16

\bibitem[\protect\citeauthoryear{Nadaraya}{Nadaraya}{1964}]{Nadaraya1964}
Nadaraya E.~A.,  1964, \mn@doi [Theory of Probability \& Its Applications]
  {10.1137/1109020}, 9, 141

\bibitem[\protect\citeauthoryear{Perlmutter et~al.,}{Perlmutter
  et~al.}{1999}]{Perlmutter1999}
Perlmutter S.,  et~al., 1999, \mn@doi [ApJ] {10.1086/307221}, 517, 565

\bibitem[\protect\citeauthoryear{Peterson et~al.,}{Peterson
  et~al.}{2021}]{Peterson2021}
Peterson E.~R.,  et~al., 2021, arXiv e-prints, \href
  {https://ui.adsabs.harvard.edu/abs/2021arXiv211003487P} {p. arXiv:2110.03487}

\bibitem[\protect\citeauthoryear{Phillips et~al.,}{Phillips
  et~al.}{2013}]{Phillips2013}
Phillips M.~M.,  et~al., 2013, \mn@doi [The Astrophysical Journal]
  {10.1088/0004-637X/779/1/38}, \href
  {https://ui.adsabs.harvard.edu/abs/2013ApJ...779...38P} {779, 38}

\bibitem[\protect\citeauthoryear{Popovic, Brout, Kessler  \& Scolnic}{Popovic
  et~al.}{2021a}]{Popovic2022}
Popovic B.,  Brout D.,  Kessler R.,   Scolnic D.,  2021a, arXiv e-prints, \href
  {https://ui.adsabs.harvard.edu/abs/2021arXiv211204456P} {p. arXiv:2112.04456}

\bibitem[\protect\citeauthoryear{Popovic, Brout, Kessler, Scolnic  \&
  Lu}{Popovic et~al.}{2021b}]{Popovic2021}
Popovic B.,  Brout D.,  Kessler R.,  Scolnic D.,   Lu L.,  2021b, \mn@doi [The
  Astrophysical Journal] {10.3847/1538-4357/abf14f}, 913, 49

\bibitem[\protect\citeauthoryear{Riess et~al.,}{Riess et~al.}{1998}]{Riess1998}
Riess A.~G.,  et~al., 1998, ApJ, 116, 1009

\bibitem[\protect\citeauthoryear{Riess et~al.,}{Riess et~al.}{2021}]{Riess2022}
Riess A.~G.,  et~al., 2021, arXiv:2112.04510 [astro-ph]

\bibitem[\protect\citeauthoryear{Rubin et~al.,}{Rubin et~al.}{2015}]{Rubin2015}
Rubin D.,  et~al., 2015, \mn@doi [ApJ] {10.1088/0004-637X/813/2/137}, 813, 137

\bibitem[\protect\citeauthoryear{Salvatier, Wiecki  \& Fonnesbeck}{Salvatier
  et~al.}{2016}]{pymc3}
Salvatier J.,  Wiecki T.~V.,   Fonnesbeck C.,  2016, \mn@doi [PeerJ Computer
  Science] {10.7717/peerj-cs.55.}, 2

\bibitem[\protect\citeauthoryear{Schultz \& Wiemer}{Schultz \&
  Wiemer}{1975}]{Schultz1975}
Schultz G.~V.,  Wiemer W.,  1975, Astronomy and Astrophysics, \href
  {https://ui.adsabs.harvard.edu/abs/1975A\&A....43..133S} {43, 133}

\bibitem[\protect\citeauthoryear{Scolnic \& Kessler}{Scolnic \&
  Kessler}{2016}]{Scolnic2016}
Scolnic D.,  Kessler R.,  2016, ApJL, 822, L35

\bibitem[\protect\citeauthoryear{Scolnic, Riess, Foley, Rest, Rodney, Brout  \&
  Jones}{Scolnic et~al.}{2014}]{Scolnic2014}
Scolnic D.~M.,  Riess A.~G.,  Foley R.~J.,  Rest A.,  Rodney S.~A.,  Brout
  D.~J.,   Jones D.~O.,  2014, \mn@doi [ApJ] {10.1088/0004-637X/780/1/37}, 780,
  37

\bibitem[\protect\citeauthoryear{Scolnic et~al.,}{Scolnic
  et~al.}{2018}]{Scolnic2018}
Scolnic D.~M.,  et~al., 2018, ApJ, 859, 101

\bibitem[\protect\citeauthoryear{Scolnic et~al.,}{Scolnic
  et~al.}{2020}]{Scolnic2020}
Scolnic D.,  et~al., 2020, ApJL, 896, L13

\bibitem[\protect\citeauthoryear{Scolnic et~al.,}{Scolnic
  et~al.}{2021}]{Scolnic2022}
Scolnic D.,  et~al., 2021, arXiv:2112.03863 [astro-ph]

\bibitem[\protect\citeauthoryear{Seabold \& Perktold}{Seabold \&
  Perktold}{2010}]{statsmodels2010}
Seabold S.,  Perktold J.,  2010, in 9th Python in Science Conference.

\bibitem[\protect\citeauthoryear{Sharon \& Kushnir}{Sharon \&
  Kushnir}{2022}]{Sharon2022}
Sharon A.,  Kushnir D.,  2022, \mn@doi [Monthly Notices of the Royal
  Astronomical Society] {10.1093/mnras/stab3380}, \href
  {https://ui.adsabs.harvard.edu/abs/2022MNRAS.509.5275S} {509, 5275}

\bibitem[\protect\citeauthoryear{Smith et~al.,}{Smith et~al.}{2020}]{Smith2020}
Smith M.,  et~al., 2020, MNRAS, 494, 4426

\bibitem[\protect\citeauthoryear{Sullivan et~al.,}{Sullivan
  et~al.}{2010}]{Sullivan2010}
Sullivan M.,  et~al., 2010, MNRAS, 406, 782

\bibitem[\protect\citeauthoryear{Suzuki et~al.,}{Suzuki
  et~al.}{2012}]{Suzuki2012}
Suzuki N.,  et~al., 2012, ApJ, 746, 85

\bibitem[\protect\citeauthoryear{Taylor, Lidman, Tucker, Brout, Hinton  \&
  Kessler}{Taylor et~al.}{2021}]{Taylor2021}
Taylor G.,  Lidman C.,  Tucker B.~E.,  Brout D.,  Hinton S.~R.,   Kessler R.,
  2021, \mn@doi [Monthly Notices of the Royal Astronomical Society]
  {10.1093/mnras/stab962}, \href
  {https://ui.adsabs.harvard.edu/abs/2021MNRAS.504.4111T} {504, 4111}

\bibitem[\protect\citeauthoryear{Tripp}{Tripp}{1998}]{Tripp1998}
Tripp R.,  1998, A\&A, 331, 815

\bibitem[\protect\citeauthoryear{VanderPlas}{VanderPlas}{2014}]{VanderPlas2014}
VanderPlas J.,  2014, arXiv, 1411.5018

\bibitem[\protect\citeauthoryear{Waskom et~al.,}{Waskom et~al.}{2020}]{seaborn}
Waskom M.,  et~al., 2020, \mn@doi [Zenodo] {10.5281/zenodo.592845}, \href
  {https://ui.adsabs.harvard.edu/abs/2020zndo...3767070W} {}

\bibitem[\protect\citeauthoryear{Watson}{Watson}{1964}]{Watson1964}
Watson G.~S.,  1964, Sankhy\=a: The Indian Journal of Statistics, Series A, 26,
  359

\makeatother
\end{thebibliography}

\appendix

\section{Full Description of the Broken-linear Model}\label{sec:fullmodel}

\begin{figure*}
    \centering
    \includegraphics[width=.60\textwidth]{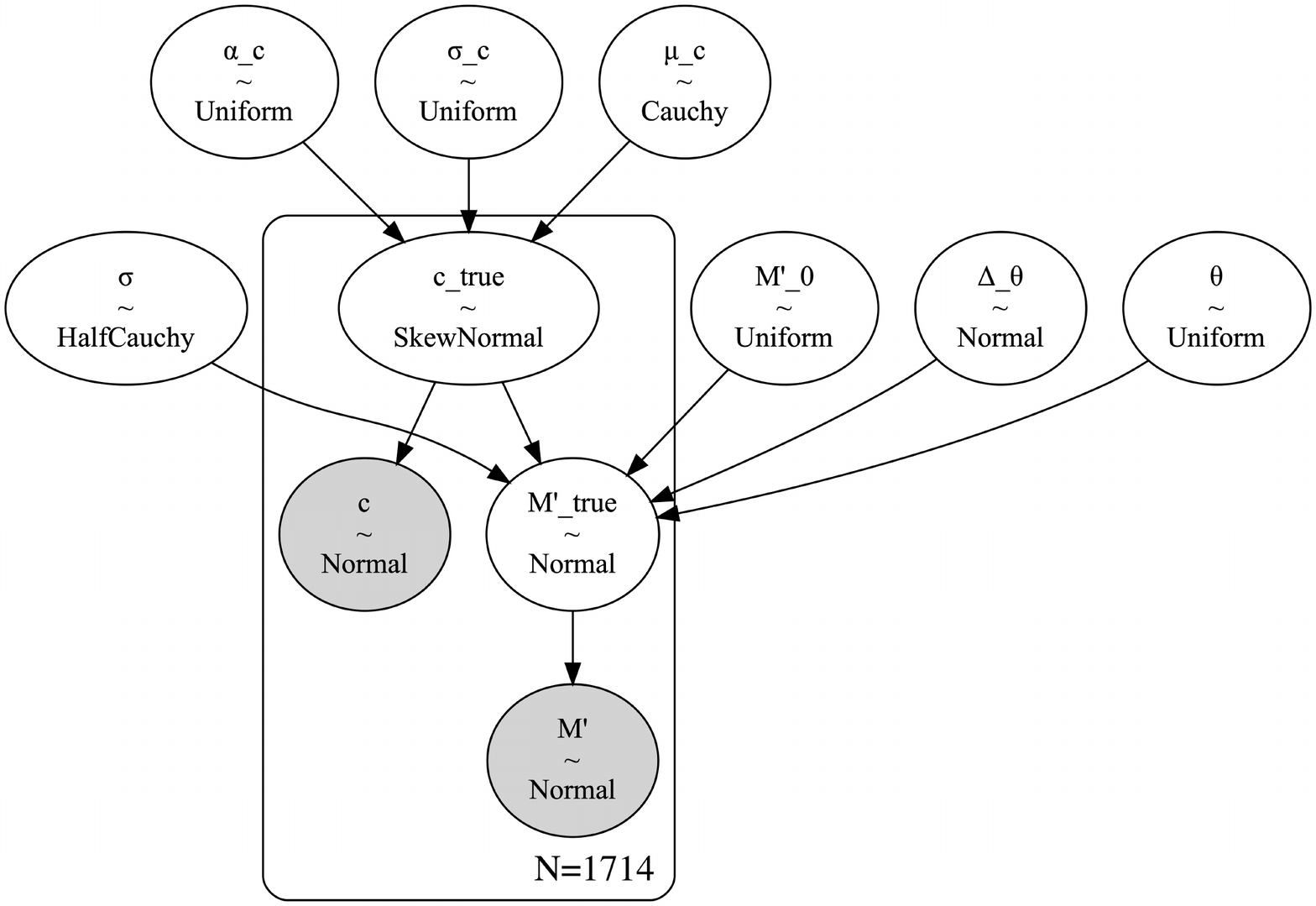}
    \caption{
    A probabilistic graphical model of our broken-linear \bhm.
    Each observed parameter (\co and $M'$) are drawn from Normal distributions with means of their true parameters (\true{c}, \true{M'}) and standard deviations of their measurement errors.
    There is one of each of these four parameters for every observed \sn ($N=\Ntotal$).
    The broken-linear relationship between \true{c} and \true{M'} is parameterized by an intercept ($M'_0$), angle of the $-0.3 \le c \le 0.3$ line ($\theta$) and change of this angle after $c> 0.3$ ($\Delta_{\theta}$).
    \true{M'} forms a Normal distribution around a mean of this broken-linear relationship, with a scatter outside the model of $\sigma$.
    The \true{c} skew-normal distribution parameters ($\mu_c$, $\sigma_c$, $\alpha_c$) are fit along with the broken-linear parameters.
    }
    \label{fig:bhm}
\end{figure*}

We built a Bayesian Hieratical model (\bhm), imitating \linmix, but using the broken-linear relationship of \cref{eqn:broken}. 
For this model, we fit the broken linear in the ``true" parameter space. These ``true'' parameters are the model's estimation of $c$ and $\mathrm{M}'$ without the effects of observational noise.
Each observed $c$ and $\mathrm{M}'$ are drawn from normal distributions ($\mathcal{N}$) with means of \true{c} and \true{\mathrm{M}'} and standard deviations equal to the measurement uncertainties ($\sigma_{\mathrm{M}'}$ and $\sigma_{c}$)
\begin{align}
    \mathrm{M}' &\sim \mathcal{N}(\true{\mathrm{M}'}, ~ \sigma_{\mathrm{M}'}^2),\\
    c &\sim \mathcal{N}(\true{c}, ~ \sigma_{c}^2).
\end{align}
The relationship between \true{c} and \true{M'} (Equation~\ref{eqn:broken}) is such that a unique unexplained scatter term ($\epsilon$) is applied to each observation. This is Equation 1 in \citet{Kelly2007}. $\epsilon$ is drawn from a normal distribution with a mean of zero and a variance of $\sigma^2$ ($\epsilon \sim \mathcal{N}(0,~ \sigma^2)$). Therefore,
\begin{equation}
    \true{\mathrm{M}'} \sim \mathcal{N}(f(\true{c}, \beta, \Delta \beta, \mathrm{M}'_0),~ \sigma^2),
\end{equation}
where the mean of each \true{\mathrm{M}'} is defined by \cref{eqn:broken}, \true{c}, and the three broken-linear parameters ($\beta$, $\Delta_{\beta}$, $\mathrm{M}'_0$).

We are able to make one significant simplification compared to \linmix. Instead of using a Gaussian mixture model to describe of the distribution of the independent variable, $c$, we are able to use a skew normal. Though population parameters have been fit in \citet{Scolnic2016} and \citet{Popovic2021}, our \bhm marginalizes over these variables. Following the parameterization of the population distribution of $c$ as a skew normal \citep{Rubin2015}, we are able to reduce the number of population parameters from six (in \linmix) to three.
Therefore, each true $c$ value (\true{c}) is drawn from a skew normal distribution where the mean, standard deviation, and skewness parameters ($\mu_c$, $s_c$, $\alpha_c$ respectively) are fit along with the broken-linear parameters to avoid biases \citep{Gull1989}.
We present our model visually in \cref{fig:bhm}.
This model differs from \linmix and \citet{Rubin2015} by
ignoring correlations between the observed parameters.

For fitting, we do a simple transformation of variables from slope ($\beta$) to angle of line above horizontal ($\theta \equiv \arctan(\beta)$). A uniform prior in slope preferentially searches larger values. As described by \citet{VanderPlas2014}, an uniformed prior is achieved with either a transform of the variables, like above, or the use of invariant Jeffreys priors  \citep{Jeffreys1946}.
When possible, we use non-informative priors, except when required for an unbiased result \citep{Gull1989}.
We present our priors in \cref{tab:priors}. 
A PyMC3 \citep{pymc3} implementation of this model and our associated analysis scripts can be found at \url{https://github.com/benjaminrose/Investigating-Red-SN}.

\begin{table*}
    \centering
    \begin{tabular}{llr}
        \hline\hline
        Variable & Name & Distribution \\
        \hline
        $\mathrm{M}'_0$ & Absolute magnitude for an $x_1=c=0$ \sn & Uniform($-19.4,~ -19.2$)\\
        $\theta$ & Slope of color-luminosity relationship & Uniform($1.2,~ 1.45$)\\
        $\Delta_{\theta}$ & Change in color-luminosity relationship slope & $\mathcal{N}(0,~ 0.15^2)$\\
        $\sigma$ & Scatter in data from outside the model & HalfCauchy($0.05$)\\
        $\true{c}$ & Estimated \sn color without measurement noise & SkewNormal($\mu_c,~ \sigma^2_c,~ \alpha_c$)\\
        $\mu_c$ & Mean of $\true{c}$ distribution & Cauchy($0,~ 0.3$)\\
        $\sigma_c$ & Standard deviation of $\true{c}$ distribution & Uniform($0.01,~ 0.2$)\\
        $\alpha_c$ & Skewness of $\true{c}$ distribution & Uniform($-0.1,~ 2.0$)\\
        \hline
    \end{tabular}
    \caption{Parameters, and their priors, used in the borken-linear \bhm.}
    \label{tab:priors}
\end{table*}

\section{Detailed Results of the Broken-linear Fit}\label{sec:fitresults}
\begin{figure*}
    \centering
    \includegraphics[width=.9\textwidth]{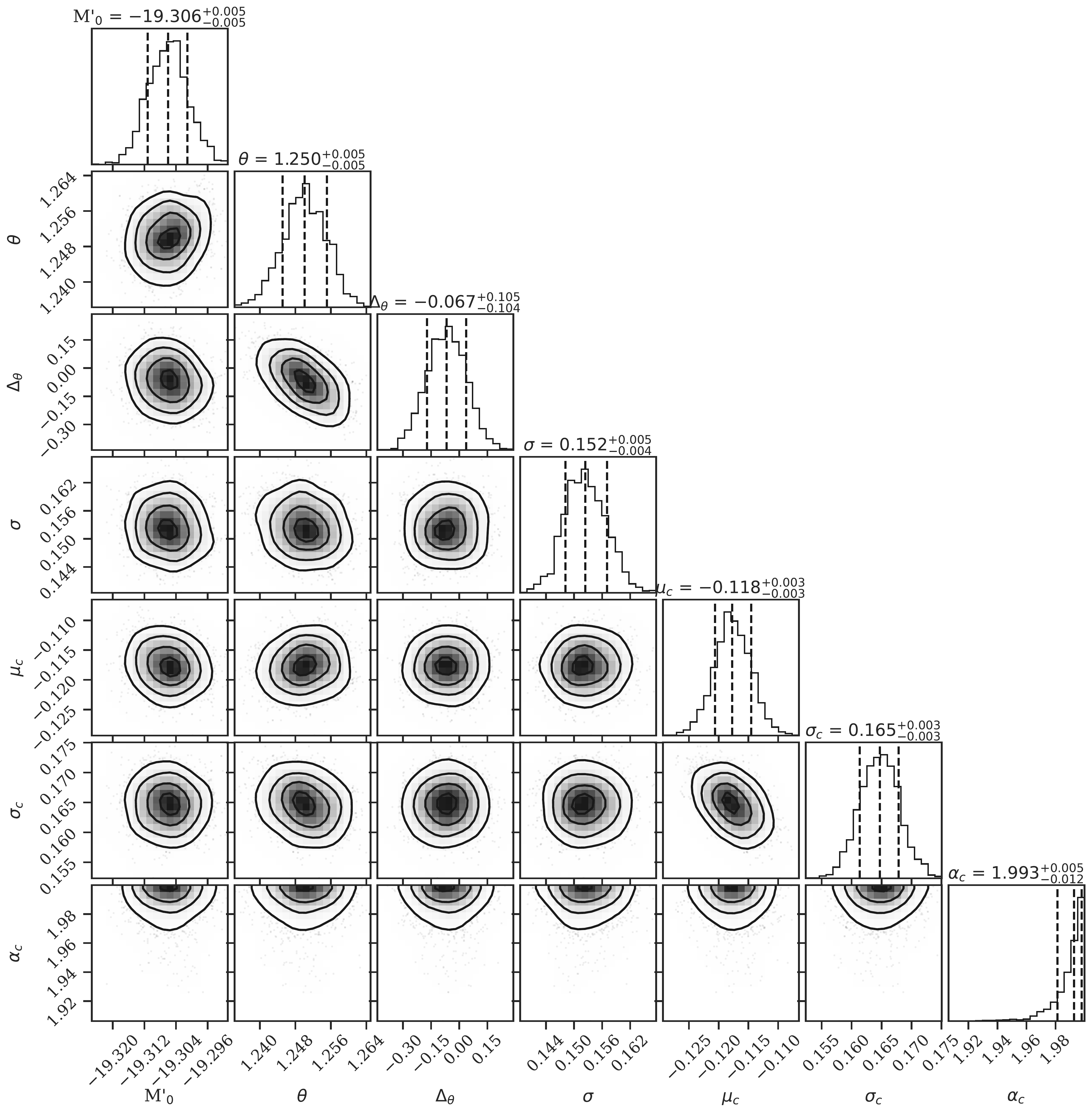}
    \caption{
    A corner plot representation of seven posterior parameters. 
    Nothing is correlated except theta and delta-theta. We do hit the edge of the colour skewness parameter, but with a good predictive posterior (\cref{fig:ppc}) we conclude that this is sufficient.
    }
    \label{fig:corner}
\end{figure*}

We present a corner plot of the model parameters in \cref{fig:corner}. All parameters are well fit ($\hat{R} = 1.0$). $\alpha_c$ hits the prior bounds. This makes sense, since we find that Gaussian tails are not a good description for the extremely red \sne. In addition, there is a strong correlation between $\theta$ and $\Delta_{\theta}$. We expect the slightly higher unexplained scatter ($\sigma = 0.19 \pm 0.01 \un{mag}$) because we do not perform a full fit, i.e., the light-curve shape standardization coefficient is fixed to $\alpha = 0.15$. 

\begin{figure*}
    \centering
    \includegraphics[width=.75\textwidth]{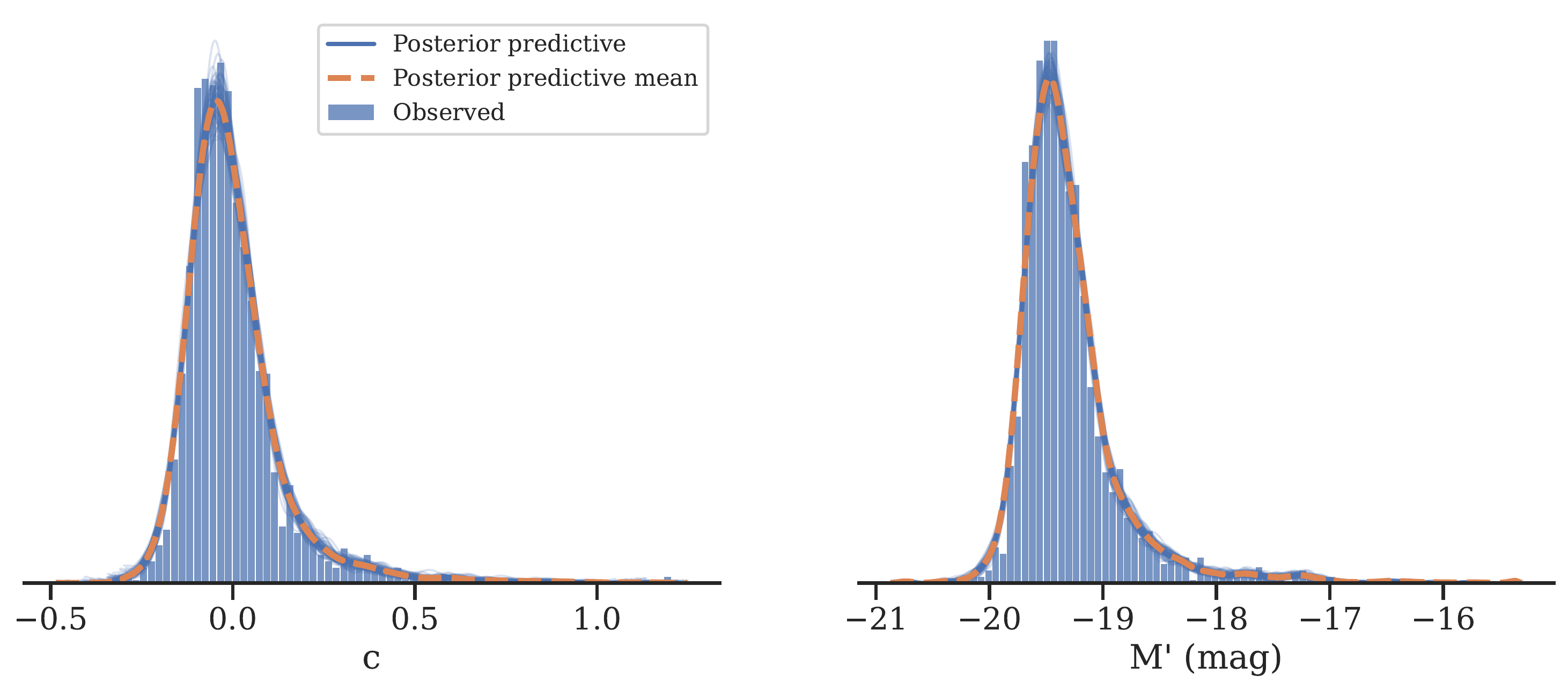}
    \caption{The posterior predictive distribution for the model described in \cref{sec:method-broken,sec:fullmodel,fig:bhm}.
    The posterior predictive distribution is the \bhm prediction of the observed parameter distributions (orange dashed line).
    This is done by marginalizing over the model parameters to get the probability of a new data set (orange dashed line), given the original data set (blue histograms).
    We see that the model is able to reproduce the observed distributions of \co (\textit{left}) and $M'$ (\textit{right}).
    }
    \label{fig:ppc}
\end{figure*}

By marginalizing over the model parameters, we are able to create a probability distribution of new data ($\vec{x}$) given the observed data ($\mathscr{X}$), $P(\vec{x}|\mathscr{X})$. This is the posterior predictive distribution.
In \cref{fig:ppc}, via the posterior predictive distribution, we see that the model can accurately recreate the observed distributions of $c$ and $M'$ implying that our parameterization of the population distributions (i.e., $\mu_c$, $s_c$, $\alpha_c$) are sufficient. 

As an additional validation, we fit the broken linear model on the \citetalias{Popovic2022} simulations directly. We successfully recovered $\beta = 3.04 \pm 0.01$ and $\Delta_{\beta} = -0.31 \pm 0.55$, for a simulation with $\beta = 3.10$ across the entire colour range.

\bsp	
\label{lastpage}
\end{document}